\documentclass[ba]{imsart}

\RequirePackage{amsthm,amsmath,amsfonts,amssymb}
\RequirePackage[authoryear]{natbib}
\RequirePackage[colorlinks,citecolor=blue,urlcolor=blue,backref=page,backref=page]{hyperref}
\RequirePackage{graphicx}

\startlocaldefs

\usepackage{graphicx}
\usepackage{multirow}
\usepackage{tikz}
\usepackage{booktabs}
\usepackage{subfigure}
\usepackage{soul}
\usepackage{lmodern}
\usepackage[boxruled, lined, longend]{algorithm2e}
\usepackage{xcolor}

\usetikzlibrary{shapes,shadows,arrows,shapes.misc}
\tikzstyle{model}=[
fill=white, text width=2em, text centered, node distance=4em]
\tikzstyle{line}=[draw,-latex']
\tikzstyle{dline}=[draw,dashed,-latex']
\tikzstyle{nline}=[draw,strike out,-latex']

\usetikzlibrary{shapes,shadows,arrows}
\tikzstyle{blank}=[text width=4em,minimum height=3em, text centered, node distance=6em]
\tikzstyle{ccyan}=[circle,fill=cyan,text width=2em,minimum height=3em, text centered, node distance=6em]
\tikzstyle{corange}=[circle,fill=orange!100,text width=2em,minimum height=3em, text centered, node distance=6em]
\tikzstyle{line}=[draw,-latex']

\newcommand{\mbf}[1]{\mathbf{#1}}
\newcommand{\mbs}[1]{\boldsymbol{#1}}
\newcommand{\mcal}[1]{\mathcal{#1}}
\newcommand{\0}{\mbs{0}}

\newcommand{\p}{{p}}
\newcommand{\pz}{{\p_0}}
\newcommand{\M}{{M}}
\newcommand{\A}{{A}}
\newcommand{\B}{{B}}

\renewcommand{\a}{{a}}

\newcommand{\MA}{{\M_\A}}
\newcommand{\MB}{{\M_\B}}

\newcommand{\pset}{\{1,\ldots,\p\}}
\newcommand{\bX}{\mbf{X}}

\newcommand{\bXA}{\mbf{X}_\A}
\newcommand{\bI}{\mbf{I}}
\newcommand{\bXz}{\mbf{X}_0}

\newcommand{\by}{\mbf{y}}
\newcommand{\bbeta}{\mbs{\beta}}
\newcommand{\bbetaA}{\mbs{\beta}_\A}
\newcommand{\bbetaz}{\mbs{\beta}_0}

\newcommand{\beps}{\mbs{\epsilon}}

\newcommand{\tauA}{\tau_\A}
\newcommand{\cC}{\mathcal{C}}
\newcommand{\cM}{\mathcal{M}}

\newcommand{\cMp}{\mathcal{M}_p}

\newcommand{\lrb}[1]{\left\{ #1\right\}}

\theoremstyle{plain}
\newtheorem{Thm}{Theorem}
\newtheorem{Prop}[Thm]{Proposition}
\newtheorem{Cor}[Thm]{Corollary}

\endlocaldefs

\pubyear{2025}
\volume{TBA}
\issue{TBA}
\firstpage{1}
\lastpage{25}        
\doi{10.1214/25-BA1580}   

\begin{document}

\begin{frontmatter}
\title{The matryoshka doll prior -- principled multiplicity correction in Bayesian model comparison}
\runtitle{The matryoshka doll model space prior}

\begin{aug}
\author[A]{\fnms{Andrew} \snm{Womack}\ead[label=e1]{womack.andrew@gmail.com}\thanksref{fca}},
\author[B]{\fnms{Daniel} \snm{Taylor-Rodr\'{i}guez}\orcid{0000-0002-2959-0281}\ead[label=e2]{dantayrod@pdx.edu}\thanksref{fca}} 
\and
\author[C]{\fnms{Claudio} \snm{Fuentes}\ead[label=e3]{fuentesc@stat.oregonstate.edu}}  
\address[A]{Unaffiliated\printead[presep={,\ }]{e1}}

\address[B]{Department of Mathematics and Statistics, Portland State University\printead[presep={,\ }]{e2}}

\address[C]{Department of Statistics, Oregon State University\printead[presep={,\ }]{e3}}
\thankstext{fca}{First coauthors}
\thanks{This is the author's accepted manuscript of an article published in
\emph{Bayesian Analysis}, Advance Publication, 2025.
DOI: \href{https://doi.org/10.1214/25-BA1580}{10.1214/25-BA1580}.}

\runauthor{A.J. Womack, D. Taylor-Rodr\'{i}guez, and C. Fuentes}
\end{aug}

\begin{abstract}
This paper introduces a general and principled construction of model space priors with a focus on regression problems. The proposed formulation regards each model as a ``local'' null hypothesis whose alternatives are the set of models that nest it. Assuming constant ratio of prior probabilities for any ``local'' null and its alternatives provides a natural isomorphism of model spaces (like a matryoshka doll), constituting an intuitive way to correct for multiplicity in Bayesian model selection and averaging problems. This isomorphism yields the Poisson distribution as the unique limiting distribution over model dimension under mild assumptions. We compare this model space prior theoretically and in simulations to widely adopted Beta-Binomial constructions. We show that the proposed prior yields a ``just-right'' multiplicity correction that induces a desirable complexity penalization profile.
\end{abstract}



\end{frontmatter}

\section{Introduction}\label{sec:intro}

The problem of identifying suitable sets of predictors in regression models of the form
\begin{equation}
\label{eq:genmodel}
\by=\bXz\bbetaz+\bX\bbeta+\beps,\quad \text{where}\quad\beps\sim N_n(\0,\bI/\tau),
\end{equation}
has been extensively studied in the statistical literature. The matrix $\bXz$ in \eqref{eq:genmodel} corresponds to $\pz$ covariates that are necessarily included in the mean structure of $\by$ (one of which must be the intercept term). The matrix $\bX$ contains the additional $\p$ covariates whose importance is not yet determined, leading to $2^\p$ possible models for the mean structure of $\by$.
We define model complexity for a model $\M$ as the number of regression coefficients $\beta_j$ in $\bbeta$ that $\M$ does not assume are $0$. This distinguishes model complexity from both model size (which includes the dimension of $\bbetaz$) as well as the rank of the model design matrix (which includes includes accounting for the columns in $\bXz$). In this manuscript, we formulate an intuitive and principled strategy to build model space prior that properly accounts for model complexity as well as the combinatorial number of models of a given complexity. 

The literature discussing priors over the model space is sparse, but important progress has been made. 
\cite{ley2009effect} and \cite{scott2010bayes} have advocated for Beta-Binomial$(a,b)$ priors on indicator variables for covariate inclusion. These priors consider the collection $\gamma_1, \ldots, \gamma_p$ of inclusion indicators ($\gamma_j = 0$ if $\beta_j = 0$ and $\gamma_j = 1$ otherwise) as independent and identically distributed (\textit{iid}) Bernoulli random variables and place a Beta$(a, b)$ prior on the probability that any given $\gamma_j$ is equal to $1$. Both \cite{ley2009effect} and \cite{scott2010bayes} recommend setting $a=b=1$, which provides a Bayesian version of a multiplicity correction. This multiplicity correction (relative to the global null model) is purely combinatorial and the probability of a model $\M$ with $k$ predictors is $P(\M) = \frac{1}{\p+1} {\p\choose k}^{-1}$.

Unfortunately, as \cite{johnson2012bayesian} note, this multiplicity correction lacks the strength to induce posterior concentration on a finite true model when local priors on the regression coefficients are used. This concentration issue arises because infinitely exchangeable Beta-Binomial priors (i.e., those with fixed $a$ and $b$) place zero mass on finite models as $\p$ increases. To correct this undesirable behavior, \cite{wilson2010bayesian} set $a=1$ and $b=\lambda\p$, which yields (a priori) a finite expected model complexity as $\p$ increases. Alternatively, \cite{castillo2015} obtain promising selection and reconstruction results by taking $b=\p^u$ for some $u>1$ and producing a faster decrease in the prior probability on model complexity. 
The Beta-Binomial priors in \cite{wilson2010bayesian} and \cite{castillo2015} must forgo infinite exchangeability of the $\gamma_j$ in order to introduce complexity penalization. 

While thinking about the model space in terms of \textit{iid} inclusion indicators with a mixing distribution on the probability of inclusion is quite natural,  this approach may lead to unintended (and perhaps undesirable) consequences.
When $a$ and $b$ are fixed, the prior on complexity from the Beta-Binomial$(a,b)$ exhibits mass loss and the probability of the set of models with $k$ predictors decreases to $0$ as $p$ increases for any fixed $k$. 
The prior on complexity from the Beta-Binomial$(a,p^u)$  for $u>1$ exhibits degeneracy and the probability of the global null model (in which every $\gamma_j$ is 0) increases to $1$ as $p$ increases. 
The Beta-Binomial$(a,\lambda p)$ produces a proper and non-degenerate prior on complexity as $p$ increases (in fact, the limit is a negative binomial distribution), sidestepping the issues of the two other priors, but has the unintended consequence of having false positive control degrade as the true model size increases  (see Section \ref{sec:comparisons}).

We believe that a better way to build a prior on the model space is by enforcing the consequences that we want (i.e., multiplicity control) and allowing the prior on model complexity to be induced by the construction.
We propose treating each model as a local null hypothesis with the set of models that nest it serving as its corresponding local alternatives.
We require that the prior probability of a model ($\M$) and the prior probability of the set of models that nest it ($\{\M^\prime:\M^\prime \text{ nests } \M\}$) are comparable. 
We enforce this comparability regardless of model complexity $k$ or total number of predictors $\p$. 
This induces a proper and non-degenerate prior on model complexity (like the Beta-Binomial$(a,\lambda p)$), but with automatic and universal false positive control as a consequence of the construction.

As a motivation for our construction, consider the Bayes' factor learning rates when we assume that we are comparing a finite ``true'' model $\M$ to another finite model $\M^\prime$ using mixtures of $g$-priors on the model-specific regression coefficients \citep[for instance][]{zellner1980posterior,berger1996intrinsic,liang2008mixtures}. If $\M$ is not nested in $\M^\prime$, then the learning rate of $\M$ versus $\M^\prime$ is exponential in $n$, which makes learning $\M$ in the pairwise comparison relatively easy. However, if $\M$ is nested in $\M^\prime$, then the Bayes' Factor learning rate of $\M$ against $\M^\prime$ is a power of $n$, which makes it difficult for $\M$ to beat $\M^\prime$ in a pairwise comparison even with large sample sizes. When considering the entire model space (and not just a pairwise comparison), the combinatorial size of the number of models nesting the true model $\M$ hinders learning. Our construction is tailored to correct for this learning issue by accounting for nesting relationships over the entire model space. 

The rest of the manuscript is structured as follows. In Section \ref{sec:localnull}, we detail the model space prior construction that is produced by using a simple proportionality principle, and discuss the intuition behind our default/recommended hyper-parameter choices. In Section \ref{sec:properties} we describe the properties of the prior, showing that the limiting distribution on model complexity is Poisson 
and demonstrating a self-similarity that motivated naming the prior the \textit{matryoshka doll}. 
In Section \ref{sec:comparisons} we compare the behavior of the matryoshka doll prior to the Beta-Binomial based prior constructions mentioned above. 
In Section \ref{sec:sims}, we empirically investigate through simulations the posterior performance of the matryoshka doll and of the Beta-Binomial constructions, in both finite model spaces and model spaces with an increasing number of covariates. 
We close the manuscript with a short discussion of ongoing research extending the proposed prior to more complex model spaces. In the online supplement, we expand the  discussion of self-similarity to generalize our construction and provide additional simulations to those presented in Section \ref{sec:sims}.

\section{Local null hypotheses and the matryoshka doll prior}
\label{sec:localnull}
The model space $\cMp$ associated with $\bX$ in \eqref{eq:genmodel} can be viewed in different ways. For instance, it has been introduced as the powerset of $\pset$ and is isomorphic to $\{0,1\}^{\times \p}$, the set of $p$-dimensional binary vectors. This binary space corresponds to the space generated by all possible binary vectors made up of inclusion indicator variables $\gamma_j$ assigned to each $j\in\pset$. In previous constructions of model space priors, the $\gamma_j$ are viewed as random variables and given a prior distribution, inducing a prior on the model space. While this seems reasonable, it requires considerable prior elicitation and specification on the model space to obtain sensible prior behavior, which is typically assessed through the ratio of prior probabilities of models or groups of models.

Here, we opt for the opposite approach, first defining a guiding principle and then building the prior following this principle. Our guiding principle is to control how each model, viewed as a local null hypothesis, relates to the set of models that nest it (i.e., its local alternative). We accomplish this by defining the prior probability of each model as a function of the prior probability of the set of models that nest it. This principle provides a consistent and straightforward multiplicity correction and induces a prior that penalizes model complexity. 

\subsection{Definition of the matryoshka doll prior}
Formally, let $\MA$ for $\A\subseteq\pset$ denote the model with mean structure defined by $\bXz\bbeta_0+\bXA\bbetaA$, where $\bXA$ is the submatrix of $\bX$ containing the columns indexed by $\a\in\A$. Model $\MA$ comes equipped with a prior distribution $\pi(\bbeta_{0},\bbetaA,\tauA|\MA)$, where $\tauA$ is the precision parameter corresponding to $\MA$. The model space $\cMp$ consists of models $\MA$ for all possible $\A\subseteq\pset$. A model $\MA$ is said to be nested in $\MB$ ($\MA\subset\MB$) whenever $\A\subsetneq\B$. We denote by $M_{\emptyset}$ the null model (i.e., where $A=\emptyset$), corresponding to the model with mean structure $\bXz\bbeta_0$. Though \eqref{eq:genmodel} corresponds to Gaussian regression, the model space prior constructed in this paper applies to regression models in general.

Given that the ratio of prior probabilities are the metric typically used to evaluate prior behavior, the simple rule we propose for building the prior is to assume that the ratio of prior probabilities for any model $\MA\in \mcal{M}_p$ (i.e., the local null hypothesis) and the set of models that nest it $\{M_C:\; C\subseteq\pset\text{ with }A\subsetneq C\}$ (i.e., the corresponding local alternative) is a fixed quantity, which we denote by $\eta$. That is, for any $A\subsetneq\pset$ we set
\begin{equation}
\label{eq:basedef}
P(\MA|\cMp,\eta)=\eta\sum_{\substack{C\subseteq\pset\\\A\subsetneq C}} P(\M_C|\cMp,\eta).
\end{equation}
The construction is completed by applying this rule uniformly over all $p$. A simple consequence of this is a straightforward calculation of the probability of the global null hypothesis provided below.

\begin{Prop}
\label{cor:pi0}
For all $p\in\mathbb{N}$ we have $P(\M_{\emptyset}|\cMp,\eta)=\eta/(1+\eta)$.
\end{Prop}
\begin{proof}
From  \eqref{eq:basedef}
$1=P(\M_{\emptyset}|\cMp,\eta)+P(\M_{\emptyset}|\cMp,\eta)/\eta$ which implies the result.
\end{proof}

Simply put,  \eqref{eq:basedef} states that the probability of a model is proportional to the sum of the probabilities of all the models that nest it, with the proportionality constant given by $\eta$. In practice, all that is needed to calculate the prior probabilities over the entire model space is selecting a value for $\eta$ (the ratio of prior probabilities of a model and all models that nest it) and using the formula recursively.  Alternative multiplicity corrections to \eqref{eq:basedef}, including one induced by controlling the ratio of prior probabilities of a model and its children models (i.e., nesting models with only one additional predictor), are discussed in Appendix A of the online supplement. 

To understand how the recursion is carried out, it helps to represent the model space as a directed acyclic graph, where the nodes correspond to models, and the edges between them define their nesting relationships (e.g., see the left graph in Figure \ref{fig:stepexample}(d)). Then, starting from the largest element in the graph (i.e., the full model), it is trivial to calculate the prior probability for each node by assigning to each of the edges of the graph a probability according to expression \eqref{eq:basedef}.

To fix ideas, consider the model space made up by the sixteen models obtained from all possible model combinations available with four predictors, with the intercept only model being the smallest model in the space. In this model space, there are five types of models determined by their complexity (i.e., models with 0, 1, 2, 3 or 4 terms). Given the symmetry of this model space, models of a particular type all have equal prior probability. As such, calculating the probability of one model per model type suffices to determine the probability distribution over the entire model space. In Figure \ref{fig:stepexample} we step through the recursive process (powered by our simple rule) used to calculate the prior over the model space, where the probability in each orange-colored node, corresponds to a proportion of the sum of the probabilities in the blue nodes that sit above them. 

Here we show the first steps of the recursion to provide the reader with some intuition for how the prior calculation following \eqref{eq:basedef} proceeds. First, assign a value to $\eta$ (see Section \ref{sec:etachoice} for some guidance), and denote by $\psi=\Pr(M_{\lrb{1,2,3,4}} | \cM_4,\eta)$ the prior probability of the full model.  Going down the graph (Figure \ref{fig:stepexample}(d)) note that model $M_{\lrb{1,2,3}}$ is only nested in $M_{\lrb{1,2,3,4}}$, hence our simple proportionality rule implies that $\Pr(M_{\lrb{1,2,3}} | \cM_4,\eta) = \eta \psi.$  Now, because $M_{\lrb{1,2}}$ is exclusively nested in models $M_{\lrb{1,2,3}}$, $M_{\lrb{1,2,4}}$, and $M_{\lrb{1,2,3,4}}$, its prior is given by $\Pr(M_{\lrb{1,2}} | \cM_4,\eta)=\eta  (\eta \psi + \eta \psi + \psi)=\eta \psi(2\eta + 1)=\eta \xi$.  The remaining steps of the recursion proceed in a similar fashion and are illustrated in Figure \ref{fig:stepexample}.  Note that computation of the prior should not use the recursion explicitly. Instead, calculations should be done using the relationships (see \eqref{eq:fact}) that stem from the prior's isomorphism, which provide superior computational performance. 

\begin{figure}
\begin{center}
{\scriptsize
\scalebox{0.7}{\subfigure[Subgraph for local null $M_{\{1,2,3\}}$]{\begin{tikzpicture}[->,>=stealth',auto,node distance=3em]
\node[blank](1_2_3){$\{1,2,3\}$};
\node[blank,above of=1_2_3](1_2_3_4){$\{1,2, 3,4\}$};
\path[line](1_2_3)--(1_2_3_4);
\end{tikzpicture}
\qquad\qquad\qquad\qquad\qquad\qquad\qquad
\begin{tikzpicture}[->,>=stealth',auto,node distance=3em]
\node[corange](1_2_3){$\eta \psi$};
\node[ccyan,above of=1_2_3](1_2_3_4){$\psi$};
\path[line](1_2_3)--(1_2_3_4);
\end{tikzpicture}}}

\bigskip

\scalebox{0.7}{\subfigure[Subgraph for local null $M_{\{1,2\}}$]{\begin{tikzpicture}[->,>=stealth',auto,node distance=3em]
\node[blank](1_2){$\{1,2\}$};
\node[blank,above of=1_2,xshift=-3em](1_2_3){$\{1,2,3\}$};
\node[blank,above of=1_2,xshift=3em](1_2_4){$\{1,2,4\}$};
\node[blank,above of=1_2,yshift=6em](1_2_3_4){$\{1,2, 3,4\}$};
\path[line](1_2)--(1_2_3);
\path[line](1_2)--(1_2_4);
\path[line](1_2_3)--(1_2_3_4);
\path[line](1_2_4)--(1_2_3_4);
\end{tikzpicture}
\qquad\qquad\qquad\qquad
\begin{tikzpicture}[->,>=stealth',auto,node distance=3em]
\node[corange](1_2){$\eta \xi$};
\node[ccyan,above of=1_2,xshift=-3em](1_2_3){$\eta \psi$};
\node[ccyan,above of=1_2,xshift=3em](1_2_4){$\eta \psi$};
\node[ccyan,above of=1_2,yshift=6em](1_2_3_4){$\psi$};
\path[line](1_2)--(1_2_3);
\path[line](1_2)--(1_2_4);
\path[line](1_2_3)--(1_2_3_4);
\path[line](1_2_4)--(1_2_3_4);
\end{tikzpicture}}}
\\
\scalebox{0.7}{\subfigure[Subgraph for local null $M_{\{1\}}$]{ \begin{tikzpicture}[->,>=stealth',auto,node distance=3em]
\node[blank](1){$\{1\}$};
\node[blank,above of=1,xshift=-6em](1_2){$\{1,2\}$};
\node[blank,right of=1_2,xshift=0em](1_3){$\{1,3\}$};
\node[blank,right of=1_3,xshift=0em](1_4){$\{1,4\}$};
\node[blank,above of=1_2](1_2_3){$\{1,2,3\}$};
\node[blank,above of=1_3](1_2_4){$\{1,2,4\}$};
\node[blank,above of=1_4](1_3_4){$\{1,3,4\}$};
\node[blank,above of=1,yshift=12em](1_2_3_4){$\{1,2, 3,4\}$};
\path[line](1)--(1_2);
\path[line](1)--(1_3);
\path[line](1)--(1_4);
\path[line](1_2)--(1_2_3);
\path[line](1_3)--(1_2_3);
\path[line](1_2)--(1_2_4);
\path[line](1_4)--(1_2_4);
\path[line](1_3)--(1_3_4);
\path[line](1_4)--(1_3_4);
\path[line](1_2_3)--(1_2_3_4);
\path[line](1_2_4)--(1_2_3_4);
\path[line](1_3_4)--(1_2_3_4);
\end{tikzpicture}
\quad\quad
\begin{tikzpicture}[->,>=stealth',auto,node distance=3em]
\node[corange](1){$\eta \kappa$};
\node[ccyan,above of=1,xshift=-6em](1_2){$\eta \xi$};
\node[ccyan,right of=1_2,xshift=0em](1_3){$\eta \xi$};
\node[ccyan,right of=1_3,xshift=0em](1_4){$\eta \xi$};
\node[ccyan,above of=1_2](1_2_3){$\eta \psi$};
\node[ccyan,above of=1_3](1_2_4){$\eta \psi$};
\node[ccyan,above of=1_4](1_3_4){$\eta \psi$};
\node[ccyan,above of=1,yshift=12em](1_2_3_4){$\psi$};
\path[line](1)--(1_2);
\path[line](1)--(1_3);
\path[line](1)--(1_4);
\path[line](1_2)--(1_2_3);
\path[line](1_3)--(1_2_3);
\path[line](1_2)--(1_2_4);
\path[line](1_4)--(1_2_4);
\path[line](1_3)--(1_3_4);
\path[line](1_4)--(1_3_4);
\path[line](1_2_3)--(1_2_3_4);
\path[line](1_2_4)--(1_2_3_4);
\path[line](1_3_4)--(1_2_3_4);
\end{tikzpicture}}}
\\
\scalebox{0.7}{\subfigure[Subgraph for local null $M_{\{\emptyset\}}$ (also represents $\cMp$)]{\begin{tikzpicture}[->,>=stealth',auto,node distance=3em]
\node[blank](null){$\emptyset$};
\node[blank,above of=null,xshift=-6em](1){$\{1\}$};
\node[blank,above of=null,xshift=-2em](2) {$\{2\}$};
\node[blank,above of=null,xshift=2em](3) {$\{3\}$};
\node[blank,above of=null,xshift=6em](4) {$\{4\}$};
\node[blank,above of=1,xshift=-4em](1_2){$\{1,2\}$};
\node[blank,right of=1_2,xshift=-1.8em](1_3){$\{1,3\}$};
\node[blank,right of=1_3,xshift=-1.8em](1_4){$\{1,4\}$};
\node[blank,right of=1_4,xshift=-1.8em](2_3){$\{2,3\}$};
\node[blank,right of=2_3,xshift=-1.8em](2_4){$\{2,4\}$};
\node[blank,right of=2_4,xshift=-1.8em](3_4){$\{3,4\}$};
\node[blank,above of=1,yshift=6em](1_2_3){$\{1,2,3\}$};
\node[blank,above of=2,yshift=6em](1_2_4){$\{1,2,4\}$};
\node[blank,above of=3,yshift=6em](1_3_4){$\{1,3,4\}$};
\node[blank,above of=4,yshift=6em](2_3_4){$\{2,3,4\}$};
\node[blank,above of=null,yshift=18em](1_2_3_4){$\{1,2, 3,4\}$};
\path[line](null)--(1);
\path[line](null)--(2);
\path[line](null)--(3);
\path[line](null)--(4);
\path[line](1)--(1_2);
\path[line](1)--(1_3);
\path[line](1)--(1_4);
\path[line](2)--(1_2);
\path[line](2)--(2_3);
\path[line](2)--(2_4);
\path[line](3)--(1_3);
\path[line](3)--(2_3);
\path[line](3)--(3_4);
\path[line](4)--(1_4);
\path[line](4)--(2_4);
\path[line](4)--(3_4);
\path[line](1_2)--(1_2_3);
\path[line](1_3)--(1_2_3);
\path[line](2_3)--(1_2_3);
\path[line](1_2)--(1_2_4);
\path[line](1_4)--(1_2_4);
\path[line](2_4)--(1_2_4);
\path[line](1_3)--(1_3_4);
\path[line](1_4)--(1_3_4);
\path[line](3_4)--(1_3_4);
\path[line](2_3)--(2_3_4);
\path[line](2_4)--(2_3_4);
\path[line](3_4)--(2_3_4);
\path[line](1_2_3)--(1_2_3_4);
\path[line](1_2_4)--(1_2_3_4);
\path[line](1_3_4)--(1_2_3_4);
\path[line](2_3_4)--(1_2_3_4);
\end{tikzpicture}
\quad\quad
\begin{tikzpicture}[->,>=stealth',auto,node distance=3em]
\node[corange](null){$\eta \delta$};
\node[ccyan,above of=null,xshift=-6em](1){$\eta \kappa$};
\node[ccyan,above of=null,xshift=-2em](2) {$\eta \kappa$};
\node[ccyan,above of=null,xshift=2em](3) {$\eta \kappa$};
\node[ccyan,above of=null,xshift=6em](4) {$\eta \kappa$};
\node[ccyan,above of=1,xshift=-4em](1_2){$\eta \xi$};
\node[ccyan,right of=1_2,xshift=-1.8em](1_3){$\eta \xi$};
\node[ccyan,right of=1_3,xshift=-1.8em](1_4){$\eta \xi$};
\node[ccyan,right of=1_4,xshift=-1.8em](2_3){$\eta \xi$};
\node[ccyan,right of=2_3,xshift=-1.8em](2_4){$\eta \xi$};
\node[ccyan,right of=2_4,xshift=-1.8em](3_4){$\eta \xi$};
\node[ccyan,above of=1,yshift=6em](1_2_3){$\eta \psi$};
\node[ccyan,above of=2,yshift=6em](1_2_4){$\eta \psi$};
\node[ccyan,above of=3,yshift=6em](1_3_4){$\eta \psi$};
\node[ccyan,above of=4,yshift=6em](2_3_4){$\eta \psi$};
\node[ccyan,above of=null,yshift=18em](1_2_3_4){$\psi$};
\path[line](null)--(1);
\path[line](null)--(2);
\path[line](null)--(3);
\path[line](null)--(4);
\path[line](1)--(1_2);
\path[line](1)--(1_3);
\path[line](1)--(1_4);
\path[line](2)--(1_2);
\path[line](2)--(2_3);
\path[line](2)--(2_4);
\path[line](3)--(1_3);
\path[line](3)--(2_3);
\path[line](3)--(3_4);
\path[line](4)--(1_4);
\path[line](4)--(2_4);
\path[line](4)--(3_4);
\path[line](1_2)--(1_2_3);
\path[line](1_3)--(1_2_3);
\path[line](2_3)--(1_2_3);
\path[line](1_2)--(1_2_4);
\path[line](1_4)--(1_2_4);
\path[line](2_4)--(1_2_4);
\path[line](1_3)--(1_3_4);
\path[line](1_4)--(1_3_4);
\path[line](3_4)--(1_3_4);
\path[line](2_3)--(2_3_4);
\path[line](2_4)--(2_3_4);
\path[line](3_4)--(2_3_4);
\path[line](1_2_3)--(1_2_3_4);
\path[line](1_2_4)--(1_2_3_4);
\path[line](1_3_4)--(1_2_3_4);
\path[line](2_3_4)--(1_2_3_4);
\end{tikzpicture}}}
}
\caption{Recursive calculation for the prior probability of one model of each type in the model space. Above $\xi=2 (\eta \psi)+\psi$, $\kappa=3(\eta \xi) + 3(\eta \psi)+\psi$, and $\delta=4 (\eta \kappa)+6(\eta \xi)+ 4(\eta \psi)+ \psi$. The probability for the full model, $\psi$, is obtained 
through normalization.}
\label{fig:stepexample}
\end{center}
\end{figure}

In general, the rule proposed in \eqref{eq:basedef} is quite simple, intuitive and only uses the structure of the model space induced by the nesting of models. The subset relation, $\subsetneq$, induces a partial ordering on $\cMp$ and the prior probability of a model is required to be proportional to the sum of the probabilities of the models that are greater than it in the partial order. Enforcing this proportionality throughout induces a multiplicity correction against local alternatives that leads to a more balanced behavior as we will evidence in later sections of the paper. The rule is also generalizable to any model space with a poset structure. For a given $\eta$, the model space prior is uniquely defined whenever the poset has a greatest element (i.e. a full model that nests all other models in $\cMp$). When the poset does not have a unique greatest element, the relative probabilities of the set of greatest elements must be also be set to define the model space prior. An example of the latter would be a regression model space with $p$ possible predictors that places an upper bound (strictly less than $p$) on the number of predictors allowed in a model. That being said, here we restrict our discussion to properties of the prior on $\cMp$ induced by \eqref{eq:basedef}.

\subsection{Choice of $\eta$}\label{sec:etachoice}

In order to meaningfully discuss the choice of $\eta$, it is useful to preview a few results from later in the paper. First, Theorem \ref{thm:poisson_lim} shows that as $p$ grows, the induced prior distribution on model complexity has a limiting Poisson distribution with mean $\theta=\log(1+1/\eta)$. Second, in Section \ref{sec:computation} we demonstrate that the convergence (of both the expectation and the distribution) is surprisingly fast and that the prior can be efficiently computed. 

These results would apparently provide straightforward means and intuition to specify $\eta$. For example, an approach to elicit a fixed $\eta$ could be through the limiting expectation $\theta=\log(1+1/\eta)$. Alternatively, the limiting Poisson distribution on model complexity could guide the specification of a prior on $\theta$ that can then be used to induce a prior on $\eta=1/(\exp(\theta)-1)$. Furthermore, the speed of prior computation would allow one to learn $\eta$ through a stochastic algorithm. These things (and more!) can all be done and should not be outside of the abilities of any Bayesian.
However, we implore the reader to eschew their Bayesian nature and not think of $\eta$ as a parameter that is to be learned. Doing so quickly leads one to accidentally negate the prior's intended multiplicity correction.

To illustrate these concerns, suppose that we have $p=1000$ predictors and \emph{a priori} we think that there should be about $10$ meaningful predictors. We could set $\theta=10$ to match this expectation, which would be equivalent to setting $\eta=1/(\exp(10)-1)\approx 4.5\times 10^{-5}$. Values for $\eta$ this small are ludicrous in terms of multiplicity correction and could produce undesirable posterior behavior. To further elaborate, assume that 10 of the $p$ predictors are strong, so we would learn them very quickly and be placed in the situation where the model containing just these 10 predictors is a local null hypothesis. In such a case, the ratio of prior probabilities of this model and the models nesting it is $4.5\times 10^{-5}$, strongly promoting the inclusion of false positives. In order to get the same posterior learning rate against a single false positive obtained with $\theta=1$, we would need to increase the sample size by a factor of $100$. 

Therefore, to avoid this type of undesirable behavior,  we urge the reader to instead simply think of $\eta$ as it was originally conceived, as the ratio of prior probabilities of any local null hypothesis against all of its alternatives. Similarly, the quantity $\theta=\log(1+1/\eta)$ should not be understood through the limiting Poisson expectation, but rather $\theta$ should be thought of as the limiting ratio of prior probabilities of the set of children models and their local null hypothesis (see Table \ref{tab:prior_odds_single} in Section \ref{sec:comp_children}). Reasonable values for $\eta$ (or $\theta$) drastically differ depending on whether one thinks of them in terms of ratios of prior probabilities or in terms of prior expected complexity.

This brings us to the specification we recommend for $\eta$. Setting $\eta>1$ (i.e, giving more weight to the local null) does help in posterior inference for learning a local null hypothesis versus its alternatives. One could even go through a laborious frequentist matching exercise (for a given sample size) using the standard ratio of posterior probabilities decision rule.  However, we advocate that the reader use either $\eta=1$ or $\theta=1$ (corresponding to a $\theta=\log(2)$ and $\eta = 1/(e-1)$, respectively) as suitable default values. They are both simple, clean, easily interpreted, and allow the structure of matryoshka doll prior to adequately express the multiplicity correction (and, in turn, the implied complexity penalization).

\section{Properties of the matryoshka doll prior}
\label{sec:properties}

In this section, we provide general properties of the prior induced by the rule given in \eqref{eq:basedef} and identify the limiting distribution of the prior.

\subsection{Finite exchangeability and conditioning isomorphism}
Before going into details, we describe the results briefly. 

First, for each $\p$ and integer $0\leq k\leq \p$, the set of models $\MA$ with $|\A|=k$ forms a complexity class and each model in the class has the same prior probability. 
For $|\A|=k\leq \p$ we can write
$
P(\MA|\cMp,\eta)=\pi_\p(k|\eta)\binom{\p}{k}^{-1}
$, where $\pi_\p(k|\eta)=P(|\A|=k|\cMp,\eta)$ is the prior on model complexity. This specification is equivalent to finite exchangeability on the inclusion indicators and is quite natural \textit{a priori}. 

Second, if we let $\B\subsetneq\pset$ and condition on models $\MA$ such that $\A\supseteq \B$, then the conditional prior $P(\MA|\A\supseteq \B,\cMp,\eta)$ behaves as if we had moved the covariates corresponding to $\B$ into $\bXz$. That is, the conditional prior is isomorphic to the prior that would have been constructed on a model space with $\p-|\B|$ covariates not in the base model. This isomorphism comes directly from the construction in \eqref{eq:basedef} and provides a number of equality constraints on the priors over complexity classes for different values of $\p$.

We now provide the formal statements and proofs. The arguments are essentially about a backwards recursion. 
We start with conditioning on $\A\supseteq\{1,\ldots,p\}$ and compute the conditional model probabilities. We then remove one covariate (for instance $p$) from required inclusion in $\A$ and condition on the event of including all but that one covariate (for instance $\A\supseteq\{1,\ldots,p-1\}$). After $k$ iterations, we have computed the conditional probabilities of the models that must contain a specific $p-k$ predictors. A recursive argument using \eqref{eq:basedef} provides a means of updating these conditional model prior probabilities when we remove an additional covariate from required inclusion in $\A$. The prior on model complexity is derived through careful book-keeping using this recursion.

\begin{Prop}
\label{prop:feprop}
Assume that \eqref{eq:basedef} holds and let $\psi_p(\eta)=P\left.\left(\M_{\pset}\right|\cMp,\eta\right)$. Then the following hold:
\begin{enumerate}
\item If $\A\subseteq \pset$ with $|\A|=\p-\ell$, then
$
P(\MA|\cMp,\eta)=f_\ell(\eta)\psi_p(\eta)
$
where $f_0(\eta)=1$ and for $\ell>0$ we have
$
f_\ell(\eta)=\eta\sum_{j=0}^{\ell-1}f_j(\eta)\binom{\ell}{j}.
$
\item Letting $\pi_\p(k|\eta)=f_{\p-k}(\eta)\psi_\p(\eta)\binom{\p}{k}$ for $k=0,\ldots,\p$, we have
$
P(\MB|\cMp,\eta)=\pi_\p(k|\eta)\binom{\p}{k}^{-1}
$
for $\B\subseteq\pset$ with $|\B|=k$.
\item For $k=0,\ldots,\p-1$, we have
$
\pi_\p(k|\eta)=\eta\sum_{j=1}^{\p-k} \pi_\p(k+j|\eta)\binom{k+j}{k}.
$
\item
$
\psi_\p(\eta)^{-1}=\sum_{j=0}^\p f_{\p-j}(\eta)\binom{\p}{j}.
$
\end{enumerate}
\end{Prop}

The first claim in Proposition \ref{prop:feprop} is proven using induction and the rest follows directly from it. The strong result from Proposition \ref{prop:feprop} is that the probability of a model of dimension $p-\ell$ only depends on $p$ through $\psi_p(\eta)$. This result allows us to make some nice equality constraints between the probabilities of sets of models of different sizes for differing numbers of potential covariates. The following corollary follows directly from Proposition \ref{prop:feprop}. 

\begin{Cor}
\label{cor:eqcor}
Let $\pi_\p(k|\eta)=f_{\p-k}(\eta)\psi_\p(\eta)\binom{\p}{k}$ for $k=0,\ldots,p$ and $p\in\mathbb{N}$.
 Let $k, \ell, \text{ and }j$ be integers satisfying $0\leq k<p$, $0<\ell\leq p-k$, and $j>0$, then
\begin{equation}
\label{eq:pieq}
\frac{\pi_\p(k+\ell|\eta)}{\pi_\p(k|\eta)}=\binom{j+k+\ell}{\ell}\binom{k+\ell}{\ell}^{-1}\frac{\pi_{\p+j}(j+k+\ell|\eta)}{\pi_{\p+j}(j+k|\eta)}.
\end{equation}
In particular, we have
\begin{equation}
\label{eq:pieq1}
\frac{\pi_\p(j+1|\eta)}{\pi_\p(j|\eta)}=\frac{1}{j+1}\frac{\pi_{\p-j}(1|\eta)}{\pi_{\p-j}(0|\eta)}
\end{equation}
for $j=0,1,2,\ldots,p-1$.
\end{Cor}

Proposition \ref{prop:feprop} also suggests the following isomorphism theorem.

\begin{Thm}
\label{thm:iso}
Fix $B\subsetneq \pset$ with $|B|=k<p$, then
\begin{equation}
P(\MA|\A\supseteq\B,\cMp,\eta)=\pi_{\p-k}(\ell|\eta)\binom{p-k}{\ell}^{-1}.
\end{equation}
for $B\subseteq A\subseteq \pset$ with $|\A|=k+\ell\leq p$.
A relative probability preserving isomorphism between the set of models  $\{\MA\in\cMp\ :\ \A\supseteq \B\}$ and the model space given by $\cM_{p-k}$ is obtained
by mapping such a set $\A$ to the set of the ranks of the elements of $\A\setminus\B$ in $\pset\setminus\B$.
\end{Thm}
\begin{proof}
The latter claim follows directly from the former. The former claim can be proven directly using the results of Proposition \ref{prop:feprop}. In particular
\begin{align*}
&P(\MA|\A\supseteq\B,\cMp,\eta)=\frac{P(\MA|\cMp,\eta)}{P(\A\supseteq\B|\cMp,\eta)}
=\frac{f_{p-k-\ell}(\eta)\psi_p(\eta)}{\sum_{j=0}^{p-k}f_{p-k-j}(\eta)\psi_p(\eta)\binom{p-k}{j}}\\
&\qquad=\frac{f_{p-k-\ell}(\eta)}{\sum_{j=0}^{p-k}f_{p-k-j}(\eta)\binom{p-k}{j}}=f_{p-k-\ell}(\eta)\psi_{p-k}(\eta)
=\pi_{\p-k}(\ell|\eta)\binom{p-k}{\ell}^{-1}.
\end{align*}
\end{proof}

The isomorphism theorem motivates naming this prior the \emph{matryoshka doll}. When one reduces the model space by deciding that some variables should be in the base model, one obtains a smaller version of the same model space. Just like opening a matryoshka doll and finding a smaller version of the same doll inside.

A partial converse to Theorem \ref{thm:iso} is true. That is, assuming a relative probability preserving isomorphism plus the assumption that $\pi_k(0|\eta)=\eta/(1+\eta)$ for all $k=1,\ldots,\p$ implies expression \eqref{eq:basedef} for a given $\p$. 
We fully characterize the priors exhibiting the isomorphism condition in Appendix A of the online supplement. 

Notice that Theorem \ref{thm:iso} provides an important difference between the matryoshka doll and infinitely exchangeable prior distributions on $\cMp$. Infinite exchangeability arises from an exchangeability and marginalization condition. In essence, we have to envision a universe with an infinite set of potential covariates to include in our models and make our inference on the proposed set of covariates by marginalizing out all but a finite number of the infinite covariates. In contrast, the matryoshka doll asks us to live in a world where the following two reasoners (with the same finite set of $p$ covariates) would reach the same conclusions: \emph{reasoner one} who decides \emph{a priori} to make $M_\B$ the base model and \emph{reasoner two} who decides \emph{a posteriori} to condition the analysis on the set of models in $\cMp$ that nest $M_\B$.

\subsection{Limiting distribution as $p$ increases}
\label{sec:limdist}
One final implication of expression \eqref{eq:basedef} is that the limit of $\pi_{p}(k|\eta)$ as $p$ increases is given by $\theta^k\exp(-\theta)/k!$ where $\theta=\log(1+1/\eta)$, which provides a limiting Poisson distribution on model size. The proof relies on the results of Proposition \ref{prop:feprop} and Corollary \ref{cor:eqcor}.  

\begin{Thm}
\label{thm:poisson_lim}
Suppose that expression \eqref{eq:basedef} holds, then
\begin{equation}
\lim_{p\rightarrow\infty} \pi_p(k|\eta) = \frac{\theta^k\exp(-\theta)}{k!}
\end{equation}
where $\theta=\log(1+1/\eta)$.
\end{Thm}
\begin{proof}
First, from \eqref{eq:pieq1}, we have
\begin{equation}
\label{eq:fact}
\frac{\pi_p(k+1|\eta)}{\pi_p(0|\eta)}=\frac{1}{(k+1)!}\prod_{j=0}^{k} \frac{\pi_{p-j}(1|\eta)}{\pi_{p-j}(0|\eta)}.
\end{equation}
Letting 
$
\theta=\lim_{\ell\rightarrow\infty} \frac{\pi_\ell(1|\eta)}{\pi_\ell(0|\eta)},
$
we have
$
\lim_{p\rightarrow\infty}\frac{\pi_p(k+1|\eta)}{\pi_p(0|\eta)}=\frac{\theta^{k+1}}{(k+1)!}
$
for 
fixed $k$. We also have
\[
1+1/\eta=1+\sum_{k=0}^{p-1} \frac{\pi_p(k+1|\eta)}{\pi_p(0|\eta)}=1+\sum_{k=0}^{p-1}\frac{1}{(k+1)!}\prod_{j=0}^{k} \frac{\pi_{p-j}(1|\eta)}{\pi_{p-j}(0|\eta)}.
\]
The Bounded Convergence Theorem provides
$1+1/\eta=\sum_{j=0}^\infty \frac{\theta^j}{j!}=\exp(\theta).$
\end{proof}

In Section \ref{sec:comparisons}, we compare the limiting Poisson distribution to the limiting distributions obtained by the prior constructions recommended by \cite{scott2010bayes}, \cite{wilson2010bayesian}, and \cite{castillo2015}. 

\subsection{Prior computation}
\label{sec:computation}

Computation can be carried out using result 3 from Proposition \ref{prop:feprop}, however it is better to carry out computation using \eqref{eq:fact} from the proof of Theorem \ref{thm:poisson_lim}. Rewriting \eqref{eq:fact} as
\begin{equation}
\label{eq:fact_rewrite}
\frac{\pi_p(k+1|\eta)}{\pi_p(1|\eta)}=\frac{1}{(k+1)!}\prod_{j=1}^{k} \frac{\pi_{p-j}(1|\eta)}{\pi_{p-j}(0|\eta)}.
\end{equation}
provides a quick iterative algorithm for computing ${\pi_p(k+1|\eta)}/{\pi_p(1|\eta)}$ for $k=1,\ldots,p-1$ using the computation of the prior for model spaces with $j=1,\ldots,p-1$ predictors. The calculation is completed using the value of $\pi_p(0|\eta)$ from Corollary \ref{cor:pi0} and the fact that
\begin{equation}
\label{eq:pi1}
\pi_p(1|\eta) = \frac{1-\pi_p(0|\eta)}{1+\sum_{j=1}^{p-1}\frac{\pi_p(k+1|\eta)}{\pi_p(1|\eta)}}.
\end{equation}

\begin{figure}
\includegraphics[trim={0 0.5cm 0 2cm}, clip, scale=0.5]{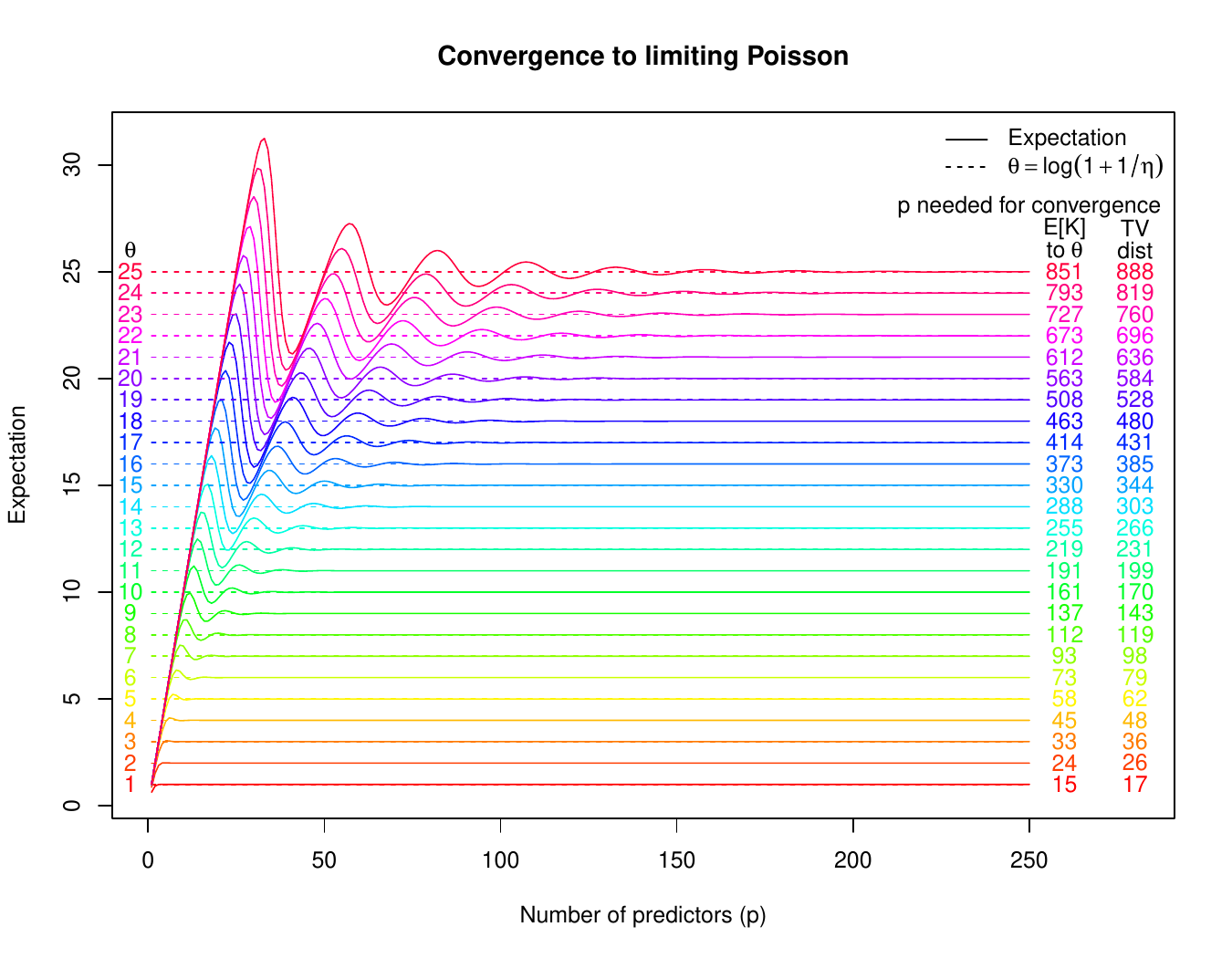}
\caption{\label{fig:expectation} The curves show convergence of the expectation $E[K_p|\eta]$ to the limiting Poisson expectation for $\theta=1,\ldots,25$ where $\eta=1/(\exp(\theta)-1)$. 
The columns on the right show the necessary number of predictors $p$ for convergence to the limiting Poisson distribution in expectation (left column) and total variation distance (right column) where convergence is determined by a distance of less than $1.82\times10^{-12}$.}
\end{figure}

Another simple consequence of \eqref{eq:fact} and \eqref{eq:fact_rewrite} is an easy computation of the prior expected model complexity.
If we let $K_p$ be the random variable on $\{0,\ldots,p\}$ representing model complexity from the matryoshka doll prior, then
\begin{equation}
\label{eq:expectation}
\mu_p(\eta) = E[K_{p}|\eta] = \frac{\pi_p(1|\eta)}{\pi_{p-1}(0|\eta)}.
\end{equation}
Though $\pi_{p}(0|\eta)$ is constant over $p$ for the matryoshka doll, we leave the expected value formula in this form because it only depends on the isomorphism from Theorem \ref{thm:iso}. Similarly, the computation of the prior using \eqref{eq:fact_rewrite} only depends on the isomorphism and the value of $\pi_p(0|\eta)$. These formulae also hold for the generalized matryoshka doll discussed in Appendix A of the online supplement.

\begin{table}[ht]
\centering
\begin{tabular}{r|rrrrrr }
  \toprule
p & min & lq & mean & median & uq & max \\ 
  \hline
100 & 0.01 & 0.02 & 0.02 & 0.02 & 0.02 & 0.05 \\ 
  500 & 0.26 & 0.29 & 0.29 & 0.29 & 0.29 & 0.97 \\ 
  1000 & 0.88 & 0.97 & 0.99 & 0.97 & 0.98 & 5.30 \\ 
  5000 & 22.12 & 24.20 & 24.58 & 24.52 & 24.82 & 48.20 \\ 
  10000 & 88.96 & 93.23 & 97.48 & 94.29 & 95.18 & 3018.28 \\ 
  50000 & 1857.47 & 1961.20 & 1976.58 & 1965.03 & 1968.45 & 4915.85 \\ 
   \bottomrule
\end{tabular}
\caption{Benchmarked computation time in milliseconds (1000 replicates) of the matryoshka doll prior in \texttt{C++}. Benchmarking done in \texttt{R} using the \texttt{Rcpp} and \texttt{microbenchmark} packages. Computations done on a Mac Mini (2024 M4 with 16GB memory).} 
\label{tab:computation_time}
\end{table}

Table \ref{tab:computation_time} shows a summary of computation times (over 1000 replicates) in milliseconds of the prior on model complexity for representative values of $p$. Once $p$ becomes too large (on the order to $10^5$), direct computation of the matryoshka doll prior does become infeasible. However, the fast convergence to the asymptotic Poisson distribution on model complexity prevents this from ever becoming an issue. For example, for one of our default values ($\theta=1$), convergence to the limiting Poisson in TV norm only requires $p= 17$. Figure \ref{fig:expectation} shows the convergence of the matryoshka to the limiting Poisson for different values of $\theta=\log(1+1/\eta)$. 

For reasonable values of $\theta$ and $p$ sufficiently large, one can shortcut the computation using \eqref{eq:fact_rewrite}, by setting $\pi_k(1|\eta)/\pi_k(0|\eta) = \theta$ for sufficiently large $k$. Alternatively, one could substitute the whole distribution on model complexity from the matryoshka doll with the (truncated) limiting Poisson distribution.

\section{Comparisons to previous constructions}
\label{sec:comparisons}

In this section, we compare the properties of the matryoshka doll prior to distributions based on widely used Beta-Binomial constructions \citep[see][]{scott2010bayes,wilson2010bayesian,castillo2015}. The prior probability of the set of models with $k$ covariates is 
\begin{equation}
\label{eq:betaprior}
\pi^{BB}_\p(k|a,b)=\frac{\Gamma(a+b)\Gamma(\p+1)\Gamma(k+a)\Gamma(\p-k+b)}{\Gamma(b)\Gamma(a)\Gamma(k+1)\Gamma(\p-k+1)\Gamma(\p+a+b)},
\end{equation}
where the superscript $BB$ denotes the Beta-Binomial prior construction. Similarly, we will use the notation $\pi^{MD}_\p(k|\eta)$ to denote the matryoshka doll prior from Section \ref{sec:localnull}. We first compare priors in terms of complexity penalization and then compare them by the associated multiplicity correction across a range of model complexities.

The matryoshka doll prior was defined with explicit multiplicity correction in mind. The complexity penalization that is obtained is a consequence of uniform application of this multiplicity correction. The real problem with the BB$(a,b)$ for fixed $b$ is mass loss. No fixed complexity class gets positive probability as $p$ increases. 
The Beta-Binomial constructions with $b$ increasing as a function of $p$ were designed to control model complexity by adding an artificially increasing number of ``failures'' into the prior.  The problem with the BB$(a, p^u)$ for $u>1$ is degeneracy. No class with positive complexity gets positive probability as $p$ increases. The comparative advantage of the BB$(a,\lambda p)$ is that it produces a proper, non-degenerate prior distribution on model complexity. 
However, as the dimension of the local null hypothesis $\M$ increases, the ratio of prior probabilities of $\M$ to its children (or ancestors) decrease to zero as a function of $|\M|$ (see Table \ref{tab:prior_odds_single} and \eqref{eq:nestratbb}). By assuming a constant probability ratio to ancestor (or children) models, the matryoshka doll construction is logically coherent avenue through which Bayesians can think about model space prior specification.

\subsection{Limiting behavior of $\pi_p(k)$}
Complexity penalization can be investigated through the ratio $\pi_p(k+1)/\pi_p(k)$, which is motivated by \cite{castillo2015}; their detection and reconstruction results rely on
\begin{equation}
\label{eq:cast}
c_1\p^{-c_2}\leq\frac{\pi_\p(k+1)}{\pi_\p(k)}\leq c_3\p^{-c_4}.
\end{equation}
 In general, for a Beta-Binomial prior, we have
\begin{equation}
\label{eq:bblimrat}
\frac{\pi^{BB}_\p(k+1|a,b)}{\pi^{BB}_\p(k|a,b)} = 
\left(\frac{k+a}{k+1}\right)\left(\frac{p-k}{p-k-1+b}\right).
\end{equation}
We are interested in the different limits for \eqref{eq:bblimrat} as $p$ increases for fixed $k$
when $b$ is a function of $p$. For ease of discussion, we will take $a=1$. 

\begin{Prop}
\label{thm:bblims}
Suppose that $k$ is finite and $a=1$ in \eqref{eq:bblimrat}, then:
\begin{enumerate}
\item If $b$ is constant as a function of $p$, then \eqref{eq:bblimrat} tends to $1$ as $p$ increases.
\item If $b=\lambda p$, then \eqref{eq:bblimrat} tends to $1/(\lambda+1)$ as $p$ increases.
\item If $b=p^u$ for $u>1$, then \eqref{eq:bblimrat} is approximated by  $p^{1-u}$ as $p$ increases.
\end{enumerate}
\end{Prop}

In more detail, Proposition \eqref{thm:bblims} implies that if $b$ is constant then the prior on model complexity exhibits mass loss; the prior probability of models with dimension less than any fixed $k$ decreases like $k/p$ as $p$ increases. Also, taking $b=\lambda p$ provides some penalization and a limiting geometric distribution on model complexity. Unfortunately, this penalization is not enough to get posterior concentration on the true model. The prior does not penalize enough to reduce false positives as the size of the true model increases, which motivated \cite{castillo2015} to require \eqref{eq:cast}. However, if $b=p^u$ for $u>1$ then the prior degenerates to a point mass on the empty model. This produces the unfortunate behavior of controlling false positives too strongly and creating false negatives for weak signals.

In contrast, the matryoshka doll prior provides
\begin{equation}
\label{eq:lnlimrat}
\lim_{p\rightarrow\infty} \frac{\pi^{MD}_\p(k+1|\eta)}{\pi^{MD}_\p(k|\eta)}
=\lim_{p\rightarrow\infty} \frac{1}{k+1}\frac{\pi_{p-k}(1|\eta)}{\pi_{p-k}(0|\eta)}
=\frac{\theta}{k+1},
\end{equation}
where $\theta=\log(1+1/\eta)$, which is an adaptive version of \eqref{eq:cast}. The limiting additional penalization for adding a covariate does not depend on the total number of covariates under consideration, rather it depends on local model complexity. When $k$ increases to be a power of $p$, then \eqref{eq:cast} holds for the matryoshka doll prior, implying a penalty adaptation that meets the requirements specified in \cite{castillo2015} without degenerating to a single point mass as $p$ increases.
In this sense, the matryoshka doll prior is a ``just right'' prior on model complexity (perhaps even Goldilocks would approve).

\subsection{Comparison to children models}\label{sec:comp_children}

In the variable selection problem the hardest models for the true model to beat are those with all of the true positives and just one false positive. These are the children models of the true model in the poset of models.  Thus, studying the prior probability ratio between a model $\M$ and a model $\M^\prime\supset M$ with $|M|+1$ predictors provides insights about the effect of the model space prior on posterior inference. 

\begin{Prop}
Let $\M$ be a model with complexity $k$ and define the children set of models to be
$\cC(\M)=\{\M^\prime\supset \M\text{ and }|\M^\prime|=|\M|+1\}$.
For $\M^\prime \in \cC(\M)$, the ratio of prior probabilities under any finitely exchangeable prior is
\begin{equation}
\label{eq:prior_odds_single}
\text{ratio}(M^\prime,M) 
= 
\frac{P(M^\prime|\cMp)}{P(M|\cMp)}
=
\frac{\pi_\p(k+1)}{\binom{p}{k+1}}\frac{\binom{p}{k}}{\pi_\p(k)}
=
\frac{k+1}{p-k}\times
\frac{\pi_\p(k+1)}{\pi_\p(k)}
\end{equation}
and the ratio of prior probabilities of the set of children models to $\M$ is
\begin{equation}
\label{eq:prodds}
\text{ratio}(\cC(M),M)\;=\;\sum_{M^\prime\in\cC(M)} \text{ratio}(M^\prime,M)\;=\;(k+1)\times\frac{\pi_p(k+1)}{\pi_p(k)}.
\end{equation}
Under the matryoshka doll, $\text{ratio}(\cC(M),M)$ is given by
\begin{equation}
\text{ratio}^{MD}(\cC(M),M\,|\,\eta) = \mu_{p-k}(\eta)\times\frac{\pi_{p-k-1}(0|\eta)}{\pi_{p-k}(0|\eta)}
\end{equation}
where $\mu_{p-k}(\eta)$ is the prior expected model complexity from \eqref{eq:expectation}. 
\end{Prop}

Table \ref{tab:prior_odds_single} displays the values for \eqref{eq:prior_odds_single} and \eqref{eq:prodds} attained by the different prior constructions and limiting situations for model complexity.  We investigate finite model size, model size growing as $\log(p)$, model size growing as $p^q$ for $0<q<1$, and model size growing as a proportion of $p$. Though this list is not exhaustive, it provides enough variation to observe the influence of the prior on posterior inference. 
There are a couple of important things to note in this table. 

First we discuss finite model size, $|\M|=k<\infty$. 
The beta-binomial with $b=p^u$ with $u>1$ provides $\text{ratio}(\cC(M),M)$ that converge to $0$ at a  $k/p^{u-1}$ rate as $p$ increases for any fixed $k$. 
The beta-binomial with constant $b$ or $b=\lambda p$ provides $\text{ratio}(\cC(M),M)$ that converge to a linear function of $k$ as $p$ increases. The slope for constant $b$ is one and the slope for $b=\lambda p$ is less than one.
The matryoshka doll provides $\text{ratio}(\cC(M),M)=\mu_{p-k}$, which converges to $\theta$ as $p$ increases.

Second we discuss increasing model size. The beta-binomial probability ratio follow the same patterns as before for each regime for $b$. Essentially, the $k$ for finite models is replaced with the appropriate growth rate for $k$ as a function of $p$.  As $|\M|$ increases with $p$, this can cause issues with false positives for the beta-binomial with constant $b$ or $b=\lambda p$, and issues with false negatives for $b=p^u$ for $u>1$.
Once again, the 
matryoshka doll provides $\text{ratio}(\cC(M),M)=\mu_{p-k}$, which converges to $\theta$ as $p$ increases so long as $p-k$ also increases to infinity.
Thus, the matryoshka doll provides nearly constant prior probability ratio for the set of children models versus $\M$ regardless of the growth rate of $k$ with $p$ (up until the case when $p-k$ is bounded above as $p$ increases).
Interestingly, the beta-binomial with $b=p^u$ requires $u\geq2$ to produce the same (or more) penalization as the matryoshka doll when $k\propto p$. 

\begin{table}[htbp]
   \centering
   \begin{tabular}{@{} lc|cccc @{}} 
      \toprule
       Prior    & exact ratio & $k$ finite & $\lim \frac{k}{\log(p)}=\xi$ & $\lim \frac{k}{p^q}=\xi$ & $\lim \frac{k}{p}=\xi$ \\
    \midrule
      \multicolumn{6}{c}{Limiting prior probability ratio of a single child model versus a model}\\
      \midrule
      $MD(\theta)$      
      & $\frac{\mu_{p-k}}{p-k}$
      & $\frac{\theta}{p}$
      & $\frac{\theta}{p}$
      & $\frac{\theta}{p}$
      & $\frac{\theta}{p(1-\xi)}$
      \\
      $BB(a,b)$       
      & $\frac{k+a}{p-k-1+b}$ 
      & $\frac{k+a}{p}$ 
      & $\frac{\log(p)}{p}$ 
      & $\frac{\xi}{p^{1-q}}$ 
      & $\xi$ 
      \\
      $BB(a,\lambda p)$     
      & $\frac{k+a}{p-k-1+\lambda p}$  
      & $\frac{k+a}{p(1+\lambda)}$ 
      & $\frac{\xi\log(p)}{p(1+\lambda)}$ 
      & $\frac{\xi}{p^{1-q}(1+\lambda)}$ 
      & $\frac{\xi}{1+\lambda}$ 
      \\
      $BB(a,p^u)$ 
      & $\frac{k+a}{p-k-1+p^u}$ 
      & $\frac{k+a}{p^u}$ 
      & $\frac{\xi\log(p)}{p^{u}}$ 
      & $\frac{\xi}{p^{u-q}}$ 
      & $\frac{\xi}{p^{u-1}}$ 
      \\
      \midrule
            \multicolumn{6}{c}{Limiting prior probability ratio of the set of children models versus a model}\\
      \midrule
      $MD(\theta)$      
      & $\mu_{p-k}$
      & $\theta$ 
      & $\theta$ 
      & $\theta$ 
      & $\theta$ 
      \\
      $BB(a,b)$       
      & $\frac{(k+a)(p-k)}{p-k-1+b}$ 
      & $k+a$ 
      & $\xi\log(p)$ 
      & ${\xi}{p^{q}}$ 
      & $\xi(1-\xi) p$ 
      \\
      $BB(a,\lambda p)$     
      & $\frac{(k+a)(p-k)}{p-k-1+\lambda p}$  
      & $\frac{k+a}{1+\lambda}$ 
      & $\frac{\xi}{1+\lambda}\log(p)$ 
      & $\frac{\xi }{1+\lambda}p^{q}$ 
      & $\frac{\xi(1-\xi) }{1+\lambda}p$ 
      \\
      $BB(a,p^u)$ 
      & $\frac{(k+a)(p-k)}{p-k-1+p^u}$ 
      & $\frac{k+a}{p^{u-1}}$ 
      & $\frac{\xi\log(p)}{p^{u-1}}$ 
      & $\frac{\xi}{p^{u-1-q}}$ 
      & $\frac{\xi(1-\xi)}{p^{u-2}}$ 
      \\
      \bottomrule
   \end{tabular}
   \caption{\label{tab:prior_odds_single} Limiting prior probability ratio of children models versus a model with $k$ predictors under different limiting regimes for $k$ as $p$ increases. In this table, $\xi>0$ is a constant (with $\xi<1$ if $k\propto p$), $u>1$, $0<q<1$, and $\theta = \log\left(1+\frac{1}{\eta}\right)$.}
\end{table}

\begin{figure}[ht]
\centering
\includegraphics[scale=.45]{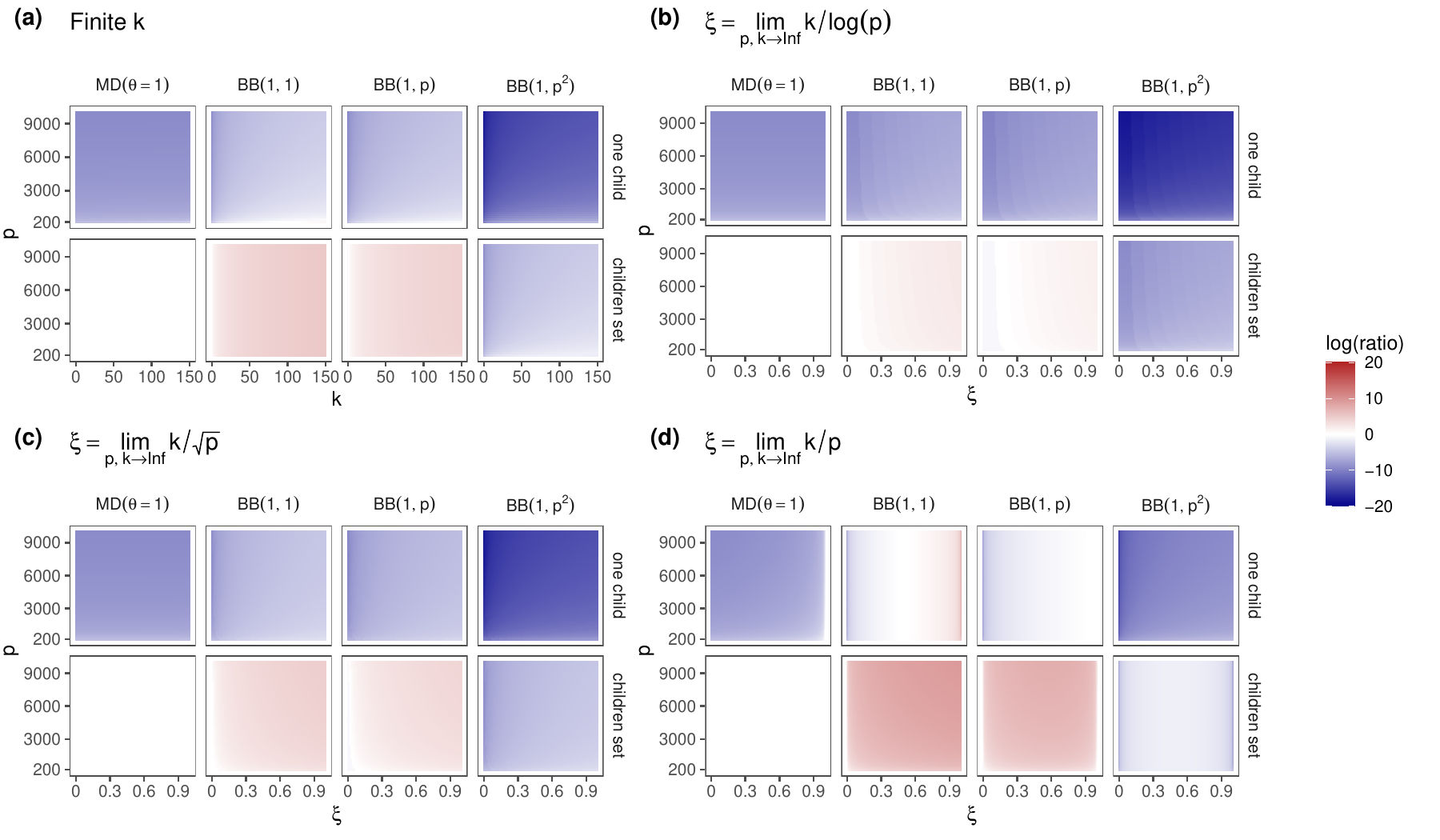} 
\caption{Comparison across model space priors in terms of the ratio of prior probabilities of a child model against a given model (i.e., \text{ratio}$(M',M)$) and for the set of children models against a model (i.e., \text{ratio}$(\cC(M),M)$) as $k$ grows with $p$ under different growth regimes, namely: (a) for finite values of  $k$, (b) as $k$ grows proportional to $\log{p}$, (c) as $k$ grows proportional to $\sqrt{p}$, and (d) as $k$ grows proportional to $p$.}
\label{fig:fms_sel}
\end{figure}

Figure \ref{fig:fms_sel} provides heatmaps of the prior probability ratio of a single child model (top row in each panel) and the set of children models (bottom row) to a given model as a function of model complexity and growth rate of $k$ with respect to $p$. For these plots, we choose $a=1$, $u=2$, and $\lambda=1$. We set $\theta=1$ so that the matryoshka doll has the same limiting expectation as the BB$(1,\lambda p)$. Some additional insights can be gleaned from the figure that are harder to understand just using Table \ref{tab:prior_odds_single}. First, the beta-binomial with either $b=1$  or $b=p$ displays an inability to properly penalize the set of children models even when $k$ is finite (panels (a), (b) and (c), bottom row). Second,  the beta-binomial with fixed $a$ and $b$ even fails to penalize a single child model when $k=\xi p$ and $\xi$ is away from 0  (panel (d), top row). Third, the beta-binomial with $b=p^2$ over-penalizes the set of children models in all scenarios except when $k=\xi p$ with $\xi$ values away from $0$ or $1$. As expected from its construction, the matryoshka doll appropriately penalizes the set of children models regardless of the size of the model or its growth rate with $p$.

\subsection{Comparison to nesting models}

The explicit multiplicity correction of the matryoshka doll construction provides constant prior probability ratio for a model versus the set of nesting models. In this subsection, we compare this to the prior probability ratio obtained through the various Beta-Binomial prior constructions. Once again, assume a model $\M$ of dimension $k$. The prior probability ratio of the models that nest $\M$ versus $\M$ is given by
\begin{equation}
\label{eq:nestrat}
\frac{\displaystyle{\sum_{M^\prime\supsetneq M}} P(M^\prime|\cMp)}{P(M|\cMp)}=\sum_{j=1}^{p-k} \frac{\pi_p(k+j)}{\pi_p(k)}\binom{k+j}{j}.
\end{equation}
Of course, because of \eqref{eq:basedef}, the matryoshka doll prior provides a value of $1/\eta$ for \eqref{eq:nestrat} regardless of $p$ and $k$. When $\pi_p(k)$ comes from the Beta-Binomial$(a,b)$ construction, then \eqref{eq:nestrat} is given by
\begin{equation}
\label{eq:nestratbb}
\frac{\displaystyle{\sum_{M^\prime\supsetneq M}} P^{BB}(M^\prime|\cMp,a,b)}{P^{BB}(M|\cMp,a,b)}
=\sum_{j=1}^{p-k} \frac{(p-k)!\ \Gamma(k+j+a)\ \Gamma(p-k-j+b)}{j!\ (p-k-j)!\ \Gamma(k+a)\ \Gamma(p-k+b)}.
\end{equation}

When $a$ and $b$ are constants, the sum in \eqref{eq:nestratbb} tends to $\infty$ as $p$ increases for any fixed $k$, exhibiting the mass loss of the prior. When $b=p^u$ for $u>1$, \eqref{eq:nestratbb} tends to $0$ as $p$ increases for any fixed $k$, showing the overly strong penalization of this prior. Further, if $u\geq 2$, then the limit of \eqref{eq:nestratbb} is $0$ even if $k$ increases with $p$ (except in the case when $k\propto p$ and $u=2$). This overly strong penalization (combined with Bayes' Factor learning rates) leads to undesired false negatives, especially of weak but meaningful predictors. 

An interesting limit occurs when $b=\lambda p$. The limit of \eqref{eq:nestratbb} for finite $k=|M|$ is
\begin{equation}
\lim_{p\rightarrow\infty}\frac{\displaystyle{\sum_{M^\prime\supsetneq M}} P^{BB}(M^\prime|\cMp,a,b=\lambda p)}{P^{BB}(M|\cMp,a,b=\lambda p)}
=
\left(1+\frac{1}{\lambda}\right)^{k+a} - 1.
\end{equation}
Thus, the prior proposed by \cite{wilson2010bayesian} does have a finite limit for each fixed $k$. However, that limit increases exponentially as a function of $k$. This explains the poor posterior concentration as the dimension of the true model increases that was discussed in \cite{castillo2015}.

The matryoshka doll strikes a balance between under- and over- penalizing the set of nesting models. Because \eqref{eq:nestrat} is a constant for all $k$ and $p$, the prior helps the Bayes' factors overcome the 
combinatorial complexity of the model space without over-penalizing. In Section \ref{sec:sims} and in the online supplement, we provide empirical evidence that constant $b$ or $b\propto p$ leads to an inflated number of false positives and lack of concentration of the posterior. Similarly, we provide empirical evidence that the degeneration of the $BB(a,p^u)$ prior to a point mass at the base model controls false positives too aggressively, leading to excessive false negatives and unwarranted posterior concentration on models that exclude covariates that explain meaningful variation in the response.

The real problem with the BB$(a,b)$ for growing $p$ is mass loss. No fixed complexity class gets positive probability as $p$ increases. The real problem with the BB$(a,p^u)$ for $u>1$ is degeneracy. No class with positive complexity gets positive probability as $p$ increases. The strength of the BB$(a,\lambda p)$ is that it produces a proper, non-degenerate prior distribution on model complexity. 
However, as the dimension of the local null hypothesis $\M$ increases, the ratio of prior probabilities of $\M$ to its children (or ancestors) decrease to zero as a function of $|\M|$ 
(see Table \ref{tab:prior_odds_single} and \eqref{eq:nestratbb}). The matryoshka doll focuses on the probability ratio as the primary quantity of prior interest. 
By assuming that the ratio of probabilities is constant to ancestor models, the matryoshka doll construction is logically coherent avenue through which Bayesians can think about model space prior specification.

\section{Performance using synthetic data sets}\label{sec:sims}

To evaluate the effects of multiplicity correction and the associated complexity penalization discussed in previous sections, we empirically assess the posterior performance of the matryoshka doll prior and compare it to that of Beta-Binomial constructions. Specifically, we consider Beta-Binomial$(a, b)$ priors with $a = 1$ and $b \in {1, p, p^2}$, alongside the matryoshka doll prior with $\theta = 1$ and $\eta = 1 / (e - 1)$. This choice of $\eta$ ensures that the matryoshka doll prior has the same asymptotic prior expectation as the Beta-Binomial$(1, p)$ prior.

We conducted three distinct sets of simulation experiments under various scenarios, including scenarios with a finite true model and others where the true model complexity  increases with sample size.  Due to space constraints, two of these simulation sets are detailed in Appendix C of the online supplement. The first supplementary set examines both increasing model complexity and varying relationships between $p$ and $n$—specifically, cases where $p < n$, $p = n$, and $p > n$. These results consistently demonstrate the inadequacy of the Beta-Binomial$(1, 1)$ and Beta-Binomial$(1, p^2)$ priors, while beginning to suggest differences between the matryoshka doll and Beta-Binomial$(1, p)$ priors when both $p$ and the size of the true model are large. The second supplementary set was designed to identify the subtle differences between the matryoshka doll prior and the Beta-Binomial$(1, p)$ prior by considering scenarios with $p > n$ and allowing both $p$ and the complexity of the true model to grow with $n$. In this section, we focus exclusively on a third set of experiments, which investigates the influence of the model space prior as either $n$ or $p$ increases, assuming a fixed sparse true model with $p \lesssim n$, which helps establish a baseline understanding of the posterior behavior induced by the different priors.

Covariates for all simulations were generated from a standard normal distribution with an assumed correlation of zero between columns. All Bayes Factors are computed relative to the global null hypothesis (the intercept-only model) using Zellner's $g$-prior (with $g=n$) on the model parameters. Zellner's $g$-prior (and all scaled mixtures of $g$-priors) provide a Bayes Factor of $1$ for models whose design matrix (including intercept column) have rank equal to the number of observations (when using singular normal distributions for the prior on the regression coefficients). We do not apply a sieve to model spaces with $p>n$ covariates. Rather, we rely on the model space prior to properly penalize model complexity. This keeps separate structural collinearities (which would arise in experimental design) and coincidental collinearities (which happen when covariates are not collinear in the population and we have $n<p$ randomly sampled observations).

For the simulations where $p$ was relatively small, we enumerated the model space. Given that enumeration with larger $p$ becomes computationally infeasible, for scenarios with $p>20$ we use an MCMC algorithm with $10^6$ MCMC iterations and $10^4$ burn-in iterations.  The proposal kernels for the MCMC are the random flip kernel, the random swap kernel, and the random model replacement kernel. The first two can be found in \cite{raftery1997bayesian} and \cite{clyde2011bayesian}, respectively. The random model replacement kernel proposes a new model with the same complexity as the current model with uniform probability and provides the possibility for much larger jumps than the random flip and random swap kernels. An R package including the code developed for the MCMC from our simulations can be found in the GitHub repository \url{https://github.com/AJWoStat/LMMCMC/}.

\subsubsection{Metrics considered}

In both sets of simulations we consider two groups of metrics. Those related to \emph{predictor selection} -- which help describe the ability of the different priors to identify good models -- and those related to \emph{posterior concentration}. All posterior averages for non-enumerable model spaces are computed using frequency based estimates of posterior probabilities from the MCMC.

\paragraph{Predictor Selection}

The main goal of any multiplicity correction is to provide a control on the family-wise false positive rate of a set of decision problems. In order to explore this in the Bayesian framework, we calculated the \emph{posterior expected true positive rates} (\textsf{tpr}), \emph{posterior expected false discovery rates} (\textsf{fdr}), and \emph{posterior expected Jaccard similarity} (\textsf{Jaccard}) under each prior for each simulation. These metrics aim to describe how well each of the priors lead the posterior distribution to concentrate on sets of \emph{good} models (i.e., those including many true predictors and few false predictors), which is essential to produce suitable model averaging results.  We also obtained the \textsf{tpr}, \textsf{fdr} and \textsf{Jaccard} of the highest probability model (HPM or mode hereforth), to assess the correspondence between the modal and the true models whenever selection is of interest.  

Letting $M_A\cap M_B$ denote the set of terms in both models $M_A$ and $M_B$, and $M_A \setminus M_B$ correspond to the set of terms in model $M_A$ but not in model $M_B$, we have that $TP_{M}=| M \cap M_T|$, $FP_{M}=| M \setminus M_T|$, and $FN_{M}=| M_T \setminus M|$ represent the true positive, false positive and false negative terms in model $M$, respectively. To approximate the distributions for the mean \textsf{tpr}, the mean \textsf{fdr}, and the mean \textsf{Jaccard} for each of the $\ell=1, 2, \ldots, 100$ data sets generated under a particular simulation scenario we calculate

\begin{eqnarray*}
\overline{\textsf{tpr}}_{\ell} &=& \sum_{M\in \cM} \textsf{tpr}(M) \cdot p(M | \by_\ell)
\;\text{with}\; \textsf{tpr}(M)=\frac{ TP_{M} }{| M_T | },\\
\overline{\textsf{fdr}}_{\ell}&=& \sum_{M\in \cM} \textsf{fdr}(M) \cdot p(M | \by_\ell)
\;\text{with}\; \textsf{fdr}(M)=\frac{ FP_{M}}{| M | },\;\text{and}\\
\overline{\textsf{Jaccard}}_{\ell}&=& \sum_{M\in \cM} \textsf{Jaccard}(M) \cdot p(M | \by_\ell)
\;\text{with}\; \textsf{Jaccard}(M)=\frac{ TP_{M} }{TP_{M} + FP_{M} + FN_{M}},
\end{eqnarray*}
where $\by_\ell$ represents the response vector generated in the $\ell$th data set. Using the 100 values obtained for these two metrics under a particular scenario, we approximate their corresponding distribution. The distributions for the \textsf{tpr}, \textsf{fdr} and  \textsf{Jaccard} for the mode are obtained by identifying the modal model $M_\ell^\star = \textsf{argmax}_{M\in\cM} p(M | \by_\ell)$ and calculating $\textsf{tpr}(M^\star_\ell)$, $\textsf{fdr}(M^\star_\ell)$ and $\textsf{Jaccard}(M^\star_\ell)$ for $\ell=1,2,\ldots,100$.  

\paragraph{Posterior Concentration}

We compute four metrics to assess posterior concentration: the posterior probability of the modal model (i.e., the HPM), the posterior probability of the true model, the number of models needed to accumulate $95\%$ of the total probability mass of the models in the space, and the rank for the true model. These four metrics help assess how each prior: (i) influences posterior mass concentration on the ``best'' model (i.e., the mode), (ii) whether or not that best model matches the true model, (iii) how much posterior mass concentration there is throughout the model space, and (iv) if the true model is contained in the set of models found, respectively.

\subsection{Simulation results}
\label{sec:simset1}

In all scenarios presented here the true model  $M_T$ is assumed to have $p_T=5$ predictors. We let $n\in\lrb{41,81, 121, 161, 201}$  for scenarios where the sample size varies, while setting the number of predictors in the full model to $p=20$.  In scenarios where we allow $p$ to grow and keep the sample size fixed, we set $n=101$ and let the number of predictors take values in $p\in\lrb{20,40,\ldots,100}$. 

To control how the signal is assigned to predictors in the true model $M_T$, we assume the $j$th largest regression coefficient in the true model, $\beta_j$, satisfies $\beta_j^2 \propto \zeta^{j}$ with $\zeta\in \lrb{1/2, 3/4, 1}$. When $\zeta=1/2$ the signal is highly concentrated on the first few predictors, whereas with $\zeta=3/4$, while still decreasing, the signal is more spread out across the true model's coefficients. We choose the value for $\sigma^2$ such that the series of coefficients produces the mean structure with signal to noise ratio $\text{snr} = \sum \beta_j^2/\sigma^2 = 2$. 

With each combination of $n$, $p$, and $\zeta$ we simulated 100 datasets to evaluate the performance of the different model-space prior specifications in terms of: (i) their variable selection ability and (ii) posterior probability mass concentration. 

Overall, as expected, increasing $n$ or reducing $p$ improved performance across all priors, with differences narrowing as data information increased. Under slower signal decay ($\zeta = 3/4$), the gap between priors diminished for tpr and Jaccard similarity; however, posterior mass allocation is over-dispersed under BB$(1,1)$, over-concentrated under BB$(1,p^2)$. The MD and BB$(1,p)$ display a more balanced behavior. Scenarios with a faster decaying signal (i.e, with $\zeta = 1/2$), especially with small $n$ or large $p$, most clearly exposed the MD prior’s advantage in maintaining balanced recovery and calibrated posterior concentration. We now present the results in some more detail.

\subsubsection{Predictor Selection}

\begin{figure}[htbp]
\centering
\subfigure[Post. mean: $p=20$, growing $n$]{\includegraphics[scale=0.37]{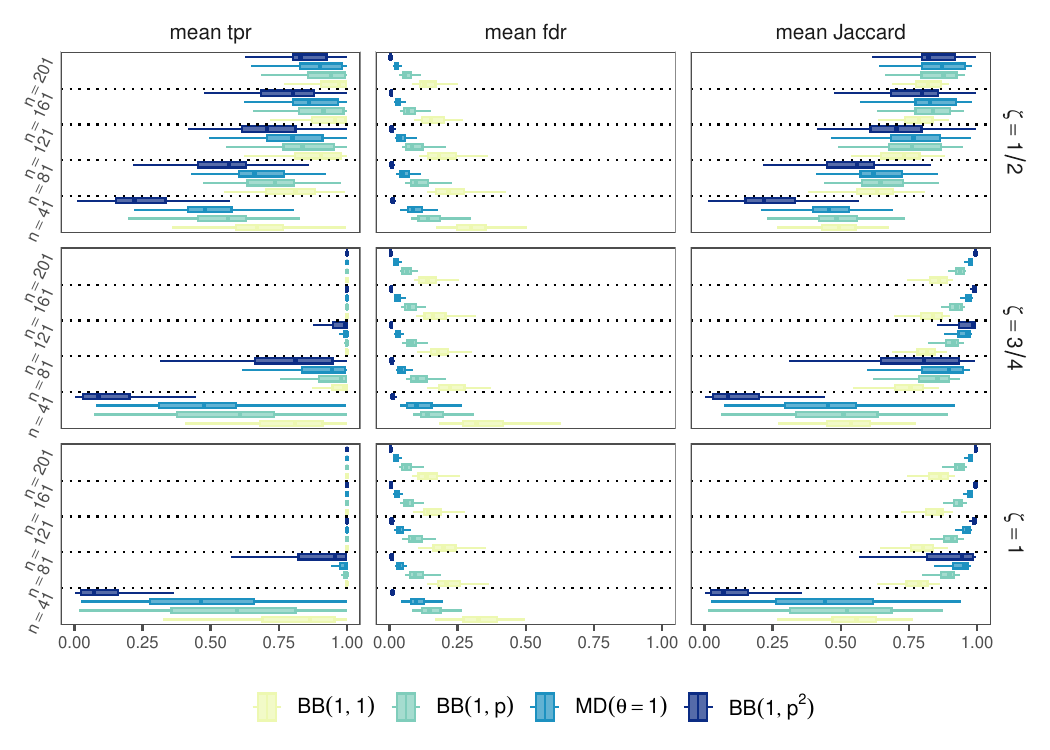}}
\subfigure[Post. mode: $p=20$, growing $n$]{\includegraphics[scale=0.37]{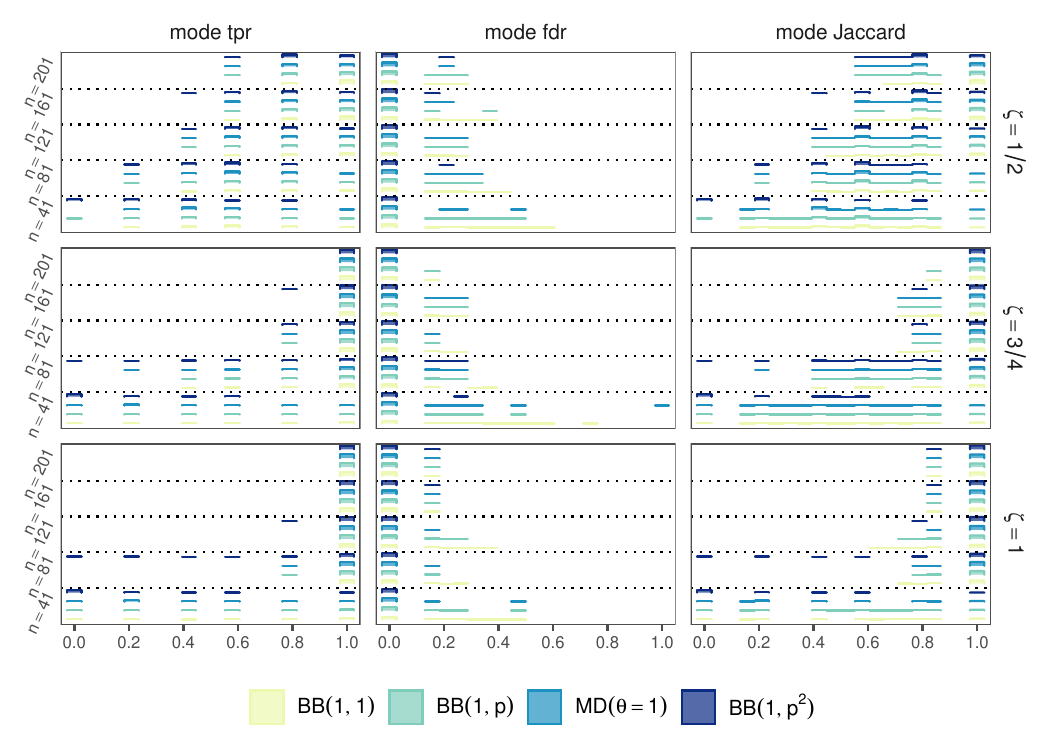}}
\subfigure[Post. mean: growing $p$, $n=101$]{\includegraphics[scale=0.37]{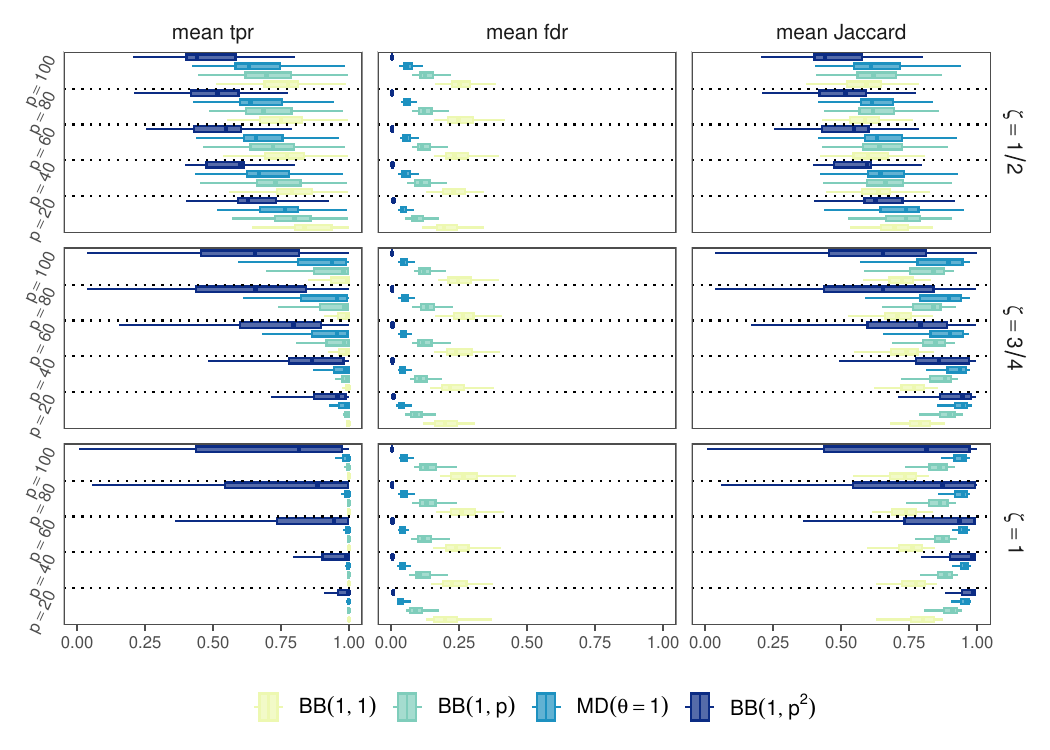}}
\subfigure[Post. mode: growing $p$, $n=101$]{\includegraphics[scale=0.37]{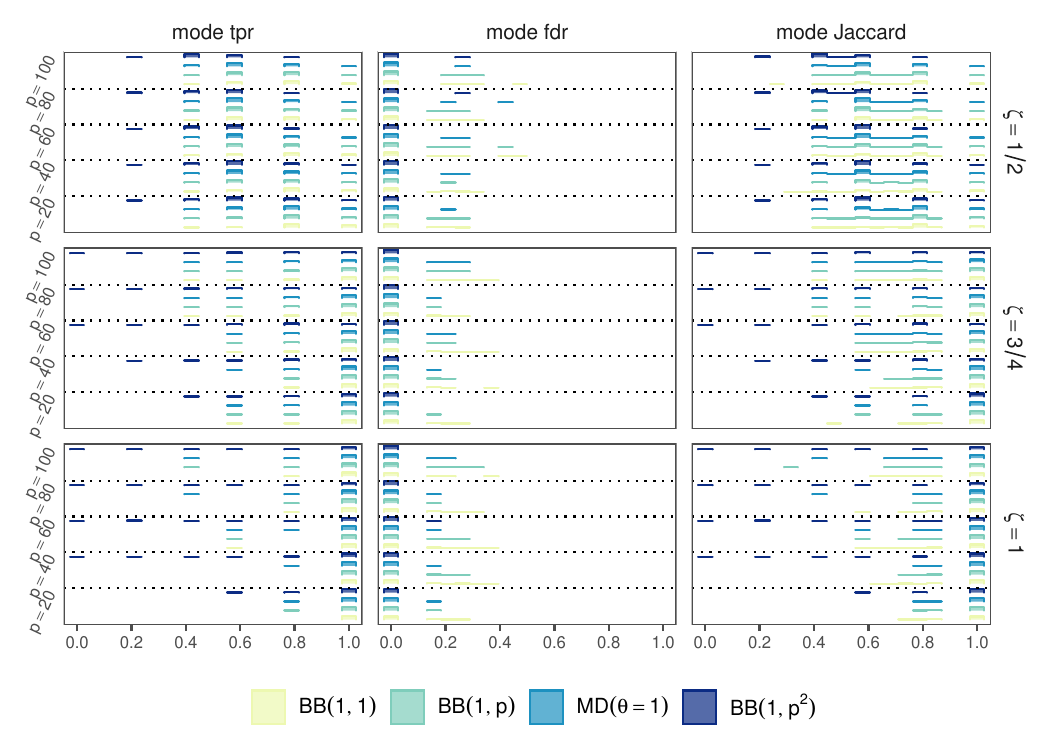}}
\caption{Expected and modal model selection metrics (TPR, FDR and Jaccard similarity) over simulation experiments in model spaces with either $n$ growing (with fixed $p=20$) or $p$ growing (with fixed $n=101$). The left column displays the boxplots for the expected values of the selection metrics obtained over the 100 replicates for each scenario by model space prior. The right column displays the distribution (as bar plots) for the different metrics of the modal model per simulation scenario.}
\label{fig:fms_pred}
\end{figure}

Across both growing $n$ and growing $p$ scenarios, MD$(\theta=1)$ provided a consistent balance between power and error control (see Figure \ref{fig:fms_pred}). Its posterior mean \textsf{tpr} was generally close to BB$(1,p)$ but with substantially lower \textsf{fdr} in large-$p$ settings, and its \textsf{fdr} was comparable to BB$(1,p^2)$ but without the sharper drop in \textsf{tpr} seen under this last prior. The MD prior also attained competitive Jaccard similarity values, indicating stable overlap between selected and true models.

The BB$(1,1)$ prior often obtained the highest \textsf{tpr} but at the expense of inflating the \textsf{fdr}, particularly under $\zeta = 1/2$, and these errors were compounded by severe posterior mass loss (see below). Conversely, BB$(1,p^2)$ produced the smallest \textsf{fdr} but systematically underfit, reducing \textsf{tpr} and \textsf{Jaccard} similarity when information was limited. BB$(1,p)$ offered more balanced results than these extremes, but in many scenarios the MD prior achieved similar or better \textsf{Jaccard} similarity with fewer false discoveries.

Results for the modal model indicate that the MD prior preserved much of the advantages mentioned above. Namely, relative to BB$(1,1)$ it achieved markedly lower \textsf{fdr} without large \textsf{tpr} losses, and compared to BB$(1,p^2)$ it avoided collapsing onto overly sparse models with low \textsf{tpr}. Differences with BB$(1,p)$ were subtler but still visible in some settings, with the MD prior producing slightly higher \textsf{Jaccard} similarity and lower \textsf{fdr} with comparable \textsf{tpr}.

\paragraph{Posterior Concentration:}

\begin{figure}[htbp]
\centering
\subfigure[$p=20$, growing $n$]{\includegraphics[scale=.5]{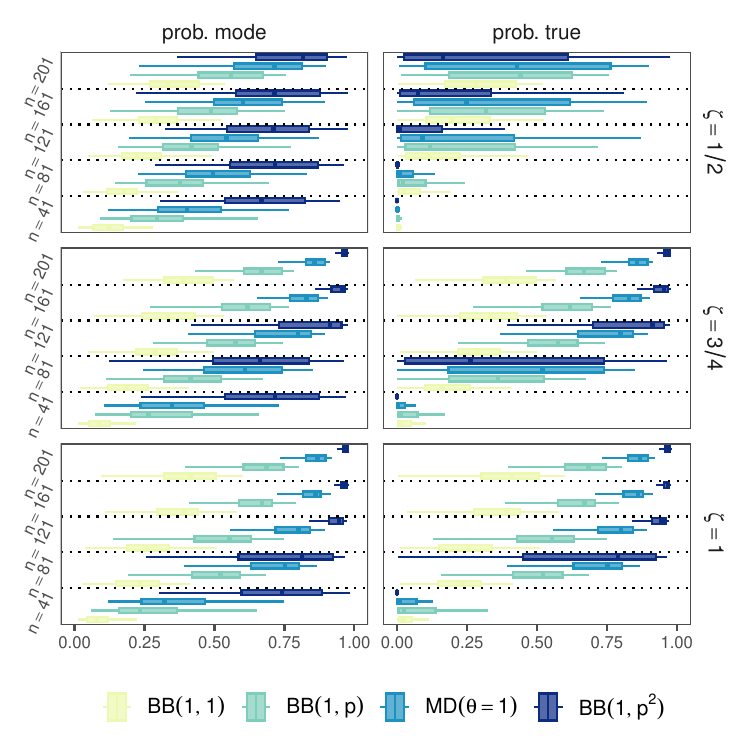} \, 
\includegraphics[scale=.5]{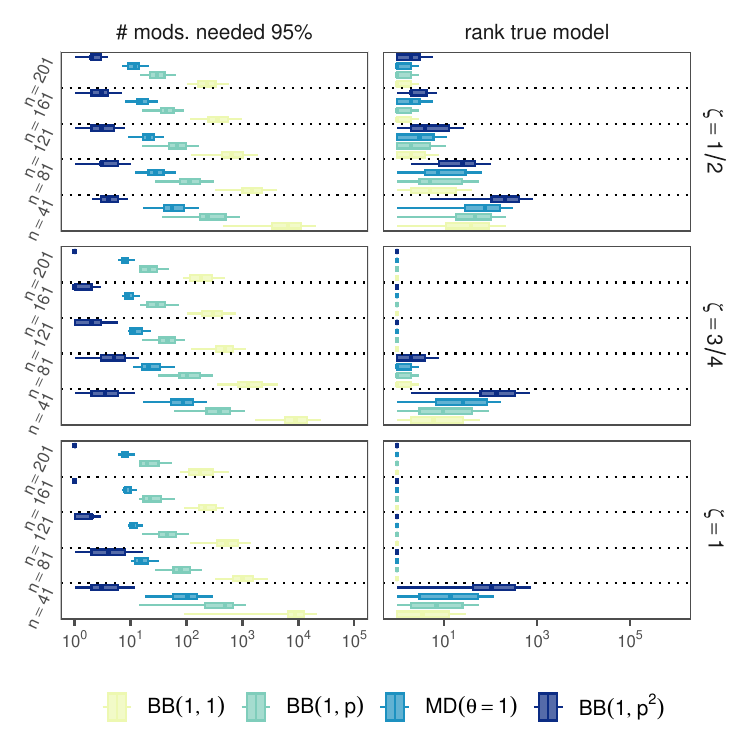}} \,
\subfigure[growing $p$, $n=101$]{\includegraphics[scale=.5]{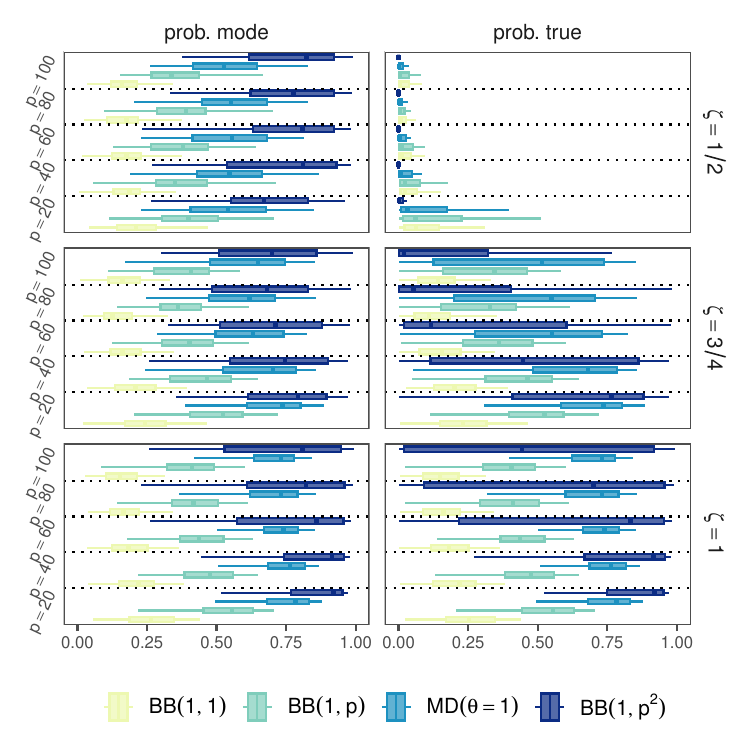} \, 
\includegraphics[scale=.5]{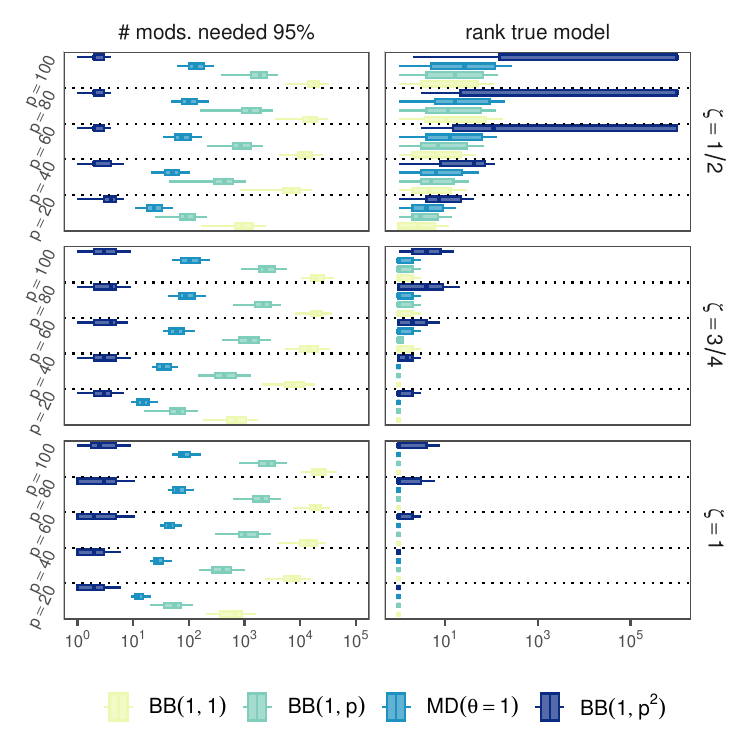}} 
\caption{Posterior concentration metrics for simulation experiments over finite model spaces for growing $n$ (fixed $p=20$) and growing $p$ (fixed $n=101$). The figure on the right displays boxplots for the number of models needed to accumulate 95\% of the posterior distribution and for the rank of the true model.}
\label{fig:fms_prob_conc}
\end{figure}

Posterior concentration metrics for this set of simulations are shown in Table \ref{tab:sim1_conc} and Figure \ref{fig:fms_prob_conc}. In Figure \ref{fig:fms_prob_conc}, the left column shows boxplots over simulations for the probability mass assigned to the modal model and the true model. In the right column, the boxplots show the number of models needed to achieve $95\%$ of the probability of the models visited by the search algorithm (with $\log_{10}$ scaled x-axis) and the rank of the true model. Note that when the true model was not found by the MCMC then its rank was set to 1,000,000 (which is the number of MCMC draws). 

The trends in these figures are striking. The BB$(1,p^2)$ produces a posterior distribution that does not properly account for model uncertainty and is generally too confident in a small set of models. It places high probability on the modal model even when it is incorrect and the maximum number of models it needed to achieve $95\%$ of the probability of the visited models mostly concentrated on less than 10 models across all scenarios. At the other end of the spectrum is the BB$(1,1)$ model, which spreads out posterior probabilities across a quite large number of models (in the thousands or ten of thousands) and places very low probability on both the modal and true models even when $p$ is small and the signal in each predictor is large.

\begin{table}[!h]
\small
\centering
\begin{tabular}[t]{l | rrr | rrr | rrr}
\toprule
\multirow{2}{*}{Prior} & \multicolumn{3}{c}{$P(\textsf{mode}|\by)$} & \multicolumn{3}{c}{$P(M_T|\by)$} &  \multicolumn{3}{c}{\# mods. needed 95\%}\\
         & mean & median & sd & mean & median & sd & mean & median & sd \\
\hline
BB$(1,1)$ & 0.225 & 0.208 & 0.122 & 0.172 & 0.147 & 0.147 & 7704 & 2982 & 9252\\
BB$(1,p)$ & 0.448 & 0.450 & 0.161 & 0.330 & 0.251 & 0.363 & 732 & 178 & 1114\\
MD$(\theta=1)$ & 0.632 & 0.672 & 0.197 & 0.453 & 0.561 & 0.350 & 52 & 30 & 58\\
BB$(1,p^2)$ & 0.768 & 0.826 & 0.197 & 0.409 & 0.192 & 0.423 & 3.2 & 2.0 & 2.4\\
\bottomrule
\end{tabular}
\caption{\label{tab:sim1_conc} Mean, median and standard deviation over all scenarios and replicates for the modal posterior probability and for the number of models needed to concentrate 95\% of the probability mass under each model space prior.}
\end{table}

The differences between the BB$(1,p^2)$, BB$(1,p)$, and MD$(\theta=1)$ priors can be easily seen in both the probability of the modal model 
and the number of models needed for $95\%$ of the probability mass of visited models (see Table \ref{tab:sim1_conc} and the disaggregated median values in Appendix B of the online supplement).  The MD$(\theta=1)$ places (on average) probability on the mode that is $1.41$ times that of the BB$(1,p)$ and $0.82$ that of the BB$(1,p^2)$.
At the same time, the BB$(1,p)$ requires an average of $14.1$ times the number of models of the MD$(\theta=1)$ for posterior concentration and the MD$(\theta=1)$ needs $16.3$ times the number of models of the BB$(1,p^2)$. While the BB$(1,p^2)$ provides far too much certainty in a small number of models (3.2 on average), the BB$(1,p)$ produces a posterior where uncertainty is spread across far too large of a set of models (732 on average). The matryoshka doll once again strikes a compromise for between these two priors and produces reasonable posterior concentration (52 on average).

\section{Discussion}\label{sec:disc}

In this paper we propose a new construction for model space priors with a specific focus on regression models. 
The key motivation behind the design of the prior is universal multiplicity correction expressed through fixing the ratio of the probabilty of a model versus the models that nest it.
This provides the prior with both a meaningful interpretation and a construction from first principles that produces a well-balanced penalization profile across model (and model space) complexity. 
The matryoshka doll prior relies on a reformulation of the way in which the model space is viewed. Instead of considering indicator variables for covariate inclusion (as is typically done in the Bayesian paradigm), the model space is viewed as a series of local null and (sets of) alternative hypotheses. This induces the isomorphism theorem and the limiting Poisson prior on model complexity presented in Section \ref{sec:properties}. 
The generalization of the matryoshka doll class of priors, which arises from the most general statement of the isomorphism condition, shows the ubiquity of the Poisson distribution on model complexity when one focuses on multiplicity correction rather than on indicator variables for covariate inclusion.

We argue that explicit multiplicity correction (and induced complexity penalization) that strikes an attractive balance that eludes beta binomial priors is a better choice for constructing model space priors.
The BB$(a,p^u)$ with $u>1$ excessively controls false positives and thus has a tendency to omit true signals.
Conversely, the BB$(a,b)$ or BB$(a,\lambda p)$ display weak control over the inclusion of false positive terms. The adaptation of the penalty to the number of covariates in the matryoshka doll provides both low penalization for adding the first few strong covariates to a model and high control of false positives when adding covariates that explain a small additional fraction of the variation in the response. 

A final property of reformulating the prior construction in terms of local null and sets of alternative hypotheses is that the explicit multiplicity control readily generalizes to other model spaces that are described by posets.
As a future direction, we are investigating the expansion of the matryoshka doll construction to more structured posets. 
In particular, we are interested in essentially non-parametric Bayesian models such as polynomial surface regression or multi-resolution analysis. The model spaces for these models can be forced to adhere to either weak or strong heredity rules \citep{taylor2016bayesian}. For a fixed maximal degree (or subspace depth), the model space is a subset of $\cMp$, but the challenge lies in accommodating heredity requirements and the notion of a complexity class. As the size of the model space grows, difficulties arise in the asymptotic and self-similarity arguments analogous to those made in Appendix A of the online supplement for computing the matryoshka doll prior on these complex model spaces.

\begin{acks}[Acknowledgments]
The authors would like to thank the anonymous referees, the Associate Editor and the Editor for extensive constructive comments that resulted in a considerably improved version of the manuscript. 
\end{acks}

\begin{funding}
Womack was partially supported by NIH award 5R01DC018813-04. Taylor-Rodr\'{i}guez was partially supported by NSF RTG DMS award 2136228, NIH award 5R01DC018813-04, and by Portland State University's FDG and Sabbatical programs. 
\end{funding}

\begin{supplement}
\stitle{\textbf{Appendix A: Generalized Matryoshka Doll}}
\sdescription{General description of priors satisfying the isomorphism condition.}
\end{supplement}

\begin{supplement}
\stitle{\textbf{Appendix B: Unpacking Table 3}}
\sdescription{Disaggregation of posterior concentration results in Table 3.}
\end{supplement}

\begin{supplement}
\stitle{\textbf{Appendix C: Simulations Sets 2 and 3}}
\sdescription{Simulation scenarios that include $p$ growing with $n$, $p>n$, and increasing size $p_T$ of the true model.}
\end{supplement}

\begin{supplement}
\stitle{\textbf{Appendix D: Additional Figures for Simulation Set 2}}
\sdescription{Appendix C.1 contains curated results for Simulation Set 2. Full figures are presented here.}
\end{supplement}

\bibliographystyle{ba}
\bibliography{wtf-2}

\end{document}